\documentclass[twocolumn]{aastex631}
\usepackage{hyperref}
\usepackage{graphicx}
\usepackage{natbib}
\usepackage{upgreek}
\usepackage{threeparttable} 
\usepackage{lipsum} 

\newcommand{\RomanNumeralCaps}[1]
    {\MakeUppercase{\romannumeral #1}}

\usepackage{xurl}
\shorttitle{L98-59\,{\rm d}'s JWST NIRSpec G395H Transmission Spectrum}
\shortauthors{A. Gressier, et al.}

\accepted{for publication in the Astrophysical Journal Letters, August 25, 2024.}

\begin{document}


\title{Hints of a sulfur-rich atmosphere around the 1.6 R$_{\oplus}$ Super-Earth L98-59\,d from JWST NIRSpec G395H transmission spectroscopy}

\author[0000-0003-0854-3002]{Am\'{e}lie Gressier}
\affiliation{Space Telescope Science Institute, 3700 San Martin Drive, Baltimore, MD 21218, USA}

\author[0000-0001-9513-1449]{N\'{e}stor Espinoza}
\affiliation{Space Telescope Science Institute, 3700 San Martin Drive, Baltimore, MD 21218, USA}
\affiliation{William H. Miller III Department of Physics and Astronomy, Johns Hopkins University, Baltimore, MD 21218, USA}

\author[0000-0002-0832-710X]{Natalie H. Allen}
\affiliation{William H. Miller III Department of Physics and Astronomy, Johns Hopkins University, Baltimore, MD 21218, USA}

\author[0000-0001-6050-7645]{David K. Sing}
\affiliation{Department of Physics and Astronomy, Johns Hopkins University, 3400 N. Charles Street, Baltimore, MD 21218, USA}
\affiliation{Department of Earth and Planetary Sciences, Johns Hopkins University, 3400 N. Charles Street, Baltimore, MD 21218, USA}

\author[0000-0002-9124-6537]{Agnibha Banerjee}
\affiliation{School of Physical Sciences, The Open University, Milton Keynes, MK7 6AA, UK}

\author[0000-0003-3726-5419]{Joanna K. Barstow}
\affiliation{School of Physical Sciences, The Open University, Milton Keynes, MK7 6AA, UK}

\author[0000-0003-3305-6281]{Jeff A. Valenti}
\affiliation{Space Telescope Science Institute, 3700 San Martin Drive, Baltimore, MD 21218, USA}

\author[0000-0002-8507-1304]{Nikole K. Lewis}
\affiliation{Department of Astronomy and Carl Sagan Institute, Cornell University, 122 Sciences Drive, Ithaca, NY 14853, USA}

\author[0000-0001-7058-1726]{Stephan M. Birkmann}
\affiliation{European Space Agency, European Space Astronomy Centre, Camino Bajo del Castillo s/n, E-28692 Villanueva de la Cañada, Madrid, Spain}

\author[0000-0002-8211-6538]{Ryan C. Challener}
\affiliation{Department of Astronomy, Cornell University, 122 Sciences Drive, Ithaca, NY 14853, USA}

\author[0000-0003-0192-6887]{Elena Manjavacas}
\affiliation{AURA for the European Space Agency (ESA), ESA Office, Space Telescope Science Institute, 3700 San Martin Drive, Baltimore, MD 21218, USA}
\affiliation{Department of Physics and Astronomy, Johns Hopkins University, 3400 N. Charles Street, Baltimore, MD 21218, USA}

\author[0000-0003-2896-4138]{Catarina Alves de Oliveira,}
\affiliation{European Space Agency, European Space Astronomy Centre, Camino Bajo del Castillo s/n, E-28692 Villanueva de la Cañada, Madrid, Spain}

\author[0000-0001-7866-8738]{Nicolas Crouzet}
\affiliation{Leiden Observatory, Leiden University, P.O. Box 9513, 2300 RA Leiden, The Netherlands}

\author[0000-0002-6881-0574]{Tracy. L Beck}
\affiliation{Space Telescope Science Institute, 3700 San Martin Drive, Baltimore, MD 21218, USA}

\begin{abstract}
Detecting atmospheres around planets with a radius below 1.6 R$_{\oplus}$, commonly referred to as rocky planets \citep{Rogers_2015, Rogers_2021}, has proven to be challenging. However, rocky planets orbiting M-dwarfs are ideal candidates due to their favorable planet-to-star radius ratio. Here, we present one transit observation of the Super-Earth L98-59\,d (1.58 R$_{\oplus}$, 2.31 M$_{\oplus}$), at the limit of rocky/gas-rich, using the JWST NIRSpec G395H mode covering the 2.8 to 5.1 $\upmu$m wavelength range. The extracted transit spectrum from a single transit observation deviates from a flat line by 2.6 to 5.6$\sigma$, depending on the data reduction and retrieval setup. The hints of an atmospheric detection are driven by a large absorption feature between 3.3 to 4.8 $\upmu$m. A stellar contamination retrieval analysis rejected the source of this feature as being due to stellar inhomogeneities, making the best fit an atmospheric model including sulfur-bearing species, suggesting that the atmosphere of L98-59\,d may not be at equilibrium. This result will need to be confirmed by the analysis of the second NIRSpec G395H visit in addition to the NIRISS SOSS transit observation.
\end{abstract}

\section{Introduction} \label{sec:intro}
Pushing the instrumental limit to possibly detect an atmosphere around an Earth-like planet is the first step in understanding whether any of these planets could support life and to what extent we could detect it. The large planet-to-star radius ratio, coupled with the high frequency of the transit event, creates favorable conditions for searching for an atmosphere using transmission spectroscopy around planets orbiting M-dwarfs. However, in the search for an atmospheric signal around a rocky planet, one must balance the M-dwarf opportunity with the possibility that long exposure to X-ray and EUV radiations can strip away the atmosphere \citep{Owen_2013, Owen_2019}. Besides, M-dwarf stars can mimic planetary atmospheric features in the transmission spectrum from photospheric inhomogeneities, known as stellar contamination \citep{Rackham_2018}. 

Space- and ground-based observations of rocky planets (R$_{\rm P}$ $<$ 1.6R$_{\oplus}$) have not yet provided strong and compelling evidence of atmospheric signals. The Hubble Space Telescope Wide Field Camera 3 Grism 141 (HST WFC3 G141), widely used in transmission spectroscopy for its water band at 1.4$\upmu$m, did not yield conclusive and/or significant evidence due to its precision and limited spectral range \citep{de_Wit_2016, de_Wit_2018, Wakeford_2019, Mugnai_2021, Edwards_2021, Gressier_2021, Garcia_2022}. Similarly, no strong constraints were found using ground-based \citep{Diamond-Lowe_2018} and Spitzer observations \citep{Demory_2011, Demory_2012, Demory_2015, Demory_2016, Kreidberg_2019, Zieba_2022, Crossfield_2022}. 

The atmospheric features in transmission for warm rocky planets are expected to reach 20 to 100 ppm, which approaches the expected noise floor of the James Webb Space Telescope instruments\citep{Greene_2016}. Even for the JWST, detecting an atmosphere around a terrestrial world is a challenge, and this has been confirmed within the first years of observations. To date, no strong evidence for an atmosphere around a rocky exoplanet has been found \citep{Lustig_Yaeger_2023, Moran_2018, May_2023, Kirk_2024}.  TRAPPIST-1\,b and c photometric data points and transmission spectra are so far consistent with no or small atmospheric signals, with the interpretation of the transmission spectrum complicated by the presence of significant stellar contamination \citep{Greene_2023, Zieba_2023, Lim_2023}. \\

The L98-59 system is composed of 3 rocky transiting planets \citep{Cloutier_2019, Kostov_2019}, with radii below 1.58 R$_{\oplus}$, and a non-transiting planet \citep{Demangeon_2021} orbiting a bright M3-dwarf (K=7.01 mag). Spanning the HST and JWST cycles 1 and 2, the L98-59 system stands out as one of the most extensively studied multi-planetary systems in emission and transmission spectroscopy (cycle 1 GO and GTO programs 1201, 1224, 2512 and cycle 2 GO 3942, 3730, 4098). So far, the HST WFC3 G141 observations in the 1.1 to 1.7$\upmu$m range have not provided strong evidence of an atmosphere for any planets in the system. Both \citet{Damiano_2022} for planet b and \citet{Zhou_2023} for planets c and d rejected cloud-free hydrogen and helium atmospheres. However, they could not rule out primary cloudy/hazy or water-rich atmospheres.\\

Here we present the result of the JWST-GTO-1224 program observations of the third planet of the system, L98-59\,d, consisting of one transit covering the 2.87 to 5.17$\upmu$m wavelength range. With a radius of 1.58$\pm$0.08R$_{\oplus}$ and a mass of 2.31$^{+0.46}_{-0.45}$M$_{\oplus}$ \citep{Luque_2022}, L98-59\,d lies on the boundary defined by \citet{Rogers_2015} between rocky and gas-rich planets, making it a highly intriguing target. In this letter, we first present the observations in Section\,\ref{sec:observations}. In Section\,\ref{sec:data_reduction}, we detail the two independent data analyses used to extract the transmission spectrum. Section\,\ref{sec:models} describes the interpretations we carried out, including atmospheric and stellar contamination forward and retrieval modeling. We discuss the results in the final Section\,\ref{sec:conclusions}.

\begin{figure*}[htpb!]
    \centering
  \includegraphics[width=\textwidth]{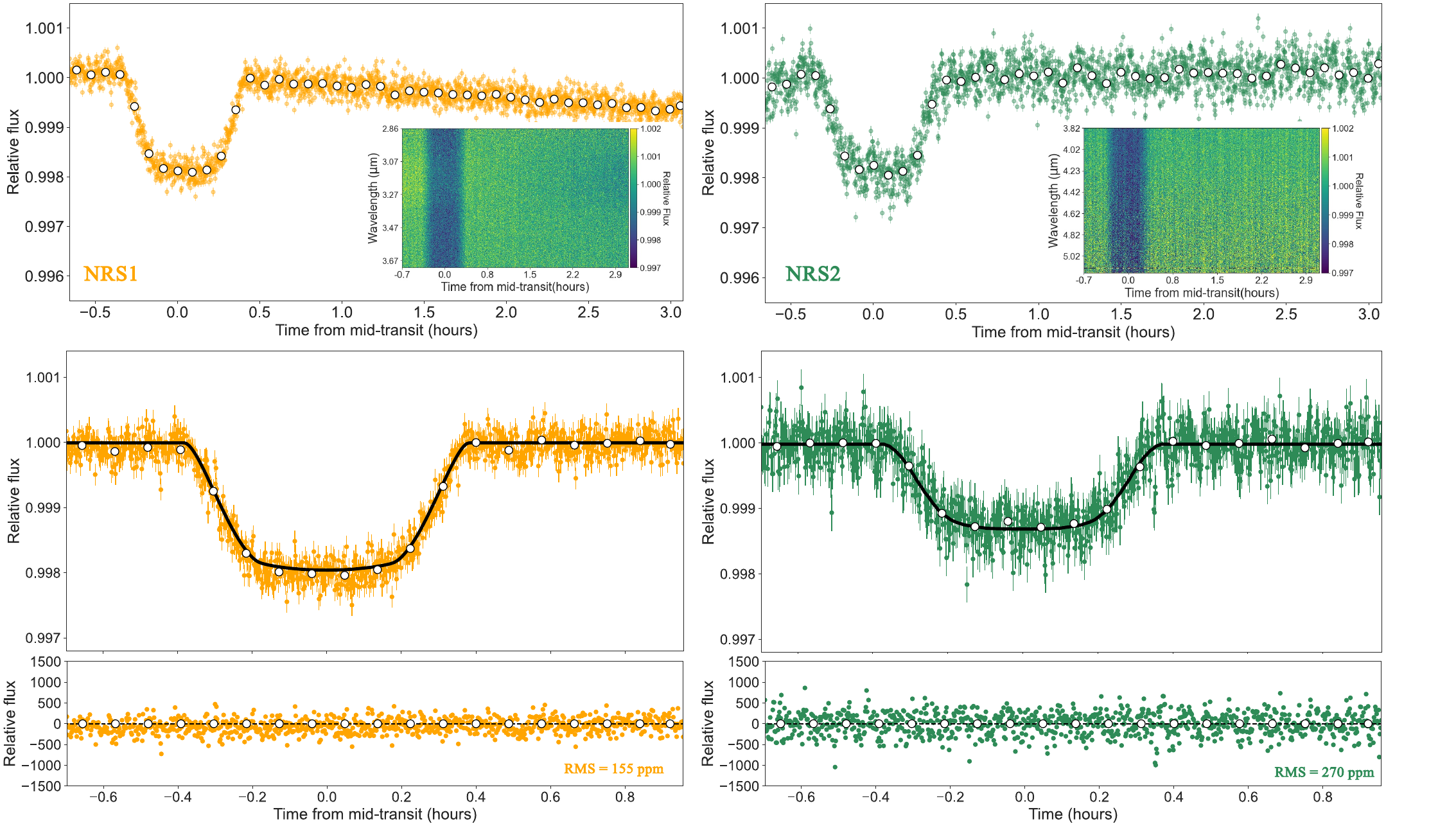}
    \caption{Top: Raw white light curves for NRS1 (left) and NRS2 (right) from \texttt{ transitspectroscopy} and spectral light curves map. Middle : Corrected white light curves (colored data points) overplotted with the best-fit transit model (black line) from \texttt{ juliet}. Bottom : residuals between raw light curves and transit model.  }
    \label{fig:wlc}
\end{figure*}
\section{JWST observations of L98-59\,\MakeLowercase{d}} \label{sec:observations}
We observed one transit of L98-59\,d using the NIRSpec/G395H mode as part of the JWST GTO cycle 1 Program 1224, led by PI Stephan Birkmann. The G395H instrument mode covers the wavelength range from 2.87 to 5.17 $\upmu$m, at a native resolution of R$\sim$2700. The grating is split over two detectors NRS1 and NRS2, with a gap between 3.72 to 3.82 $\upmu$m. The observations started on June 25$^{\rm nd}$ 2023, covered the full transit and sufficient baseline for a total of 5.34 hours, resulting in 2121 integrations, each composed of 6 groups up-the-ramp. The observation used the NIRSpec Bright Object Time Series mode with the NRSRAPID readout pattern, S1600A1 slit, and the SUB2048 subarray. 

\section{Data reduction}  \label{sec:data_reduction}
The reduction of L98-59\,d NIRSpec G395H observations was obtained using two independent pipelines, \texttt{ transitspectroscopy} \citep{espinoza_nestor_2022} and  \texttt{FIREFLy}  \citep{Rustamkulov_2022, Rustamkulov_2023} which have already been benchmarked for the JWST Early Release Science (ERS) program \citep{Alderson_2023, Rustamkulov_2023}.  \texttt{FIREFLy} have provided data reductions for smaller planets observations with the JWST NIRSpec G395H \citep{Lustig_Yaeger_2023,Moran_2023, May_2023} which makes it a good comparison. 

\subsection{Data reduction with \texttt{transitspectroscopy}} \label{sec:transitspectroscopy}
We used the \texttt{jwst} pipeline version 1.12.5 and the \texttt{transitspectroscopy} pipeline \citep{espinoza_nestor_2022}\footnote{\url{https://github.com/nespinoza/transitspectroscopy}} to reduce the NIRSpec G395H transit of L98-59\,d. The reduction process followed the  \texttt{jwst} pipeline from uncal.fits to rateints.fits, excluding the jump step detection, which we customized using the \texttt{transitspectroscopy} algorithm. This algorithm identifies outliers at the pixel level within each group by calculating group differences, applying a median filter, and detecting jumps. Spectral tracing was performed using the \texttt{trace\_spectrum} function, smoothed with a spline function. We determined the trace position using a cross-correlation method with a Gaussian input, covering specified pixel ranges for NRS1 and NRS2. Background estimation involved masking pixels around the trace and calculating the median of the remaining pixels. To address 1/f noise \citep{Jakobsen_2022, Birkmann_2022} -- noise arising from the electronics of the readout pattern, which appears as column striping in the subarray image -- we subtracted the median out-of-transit frame and iteratively removed the median of non-omitted pixels near the trace position. The stellar spectrum was extracted using the \texttt{getSimpleSpectrum} routine with a 3-pixel radius aperture, minimizing out-of-transit flux. We employed a box-extraction method to sum flux within the aperture and replaced outliers exceeding a 5$\sigma$ threshold with a 1D median filter. White light curves and pixel-level light curves were generated from the time series of 1D stellar spectra for both detectors. \\

\begin{figure*}
    \centering
  \includegraphics[width=\textwidth]{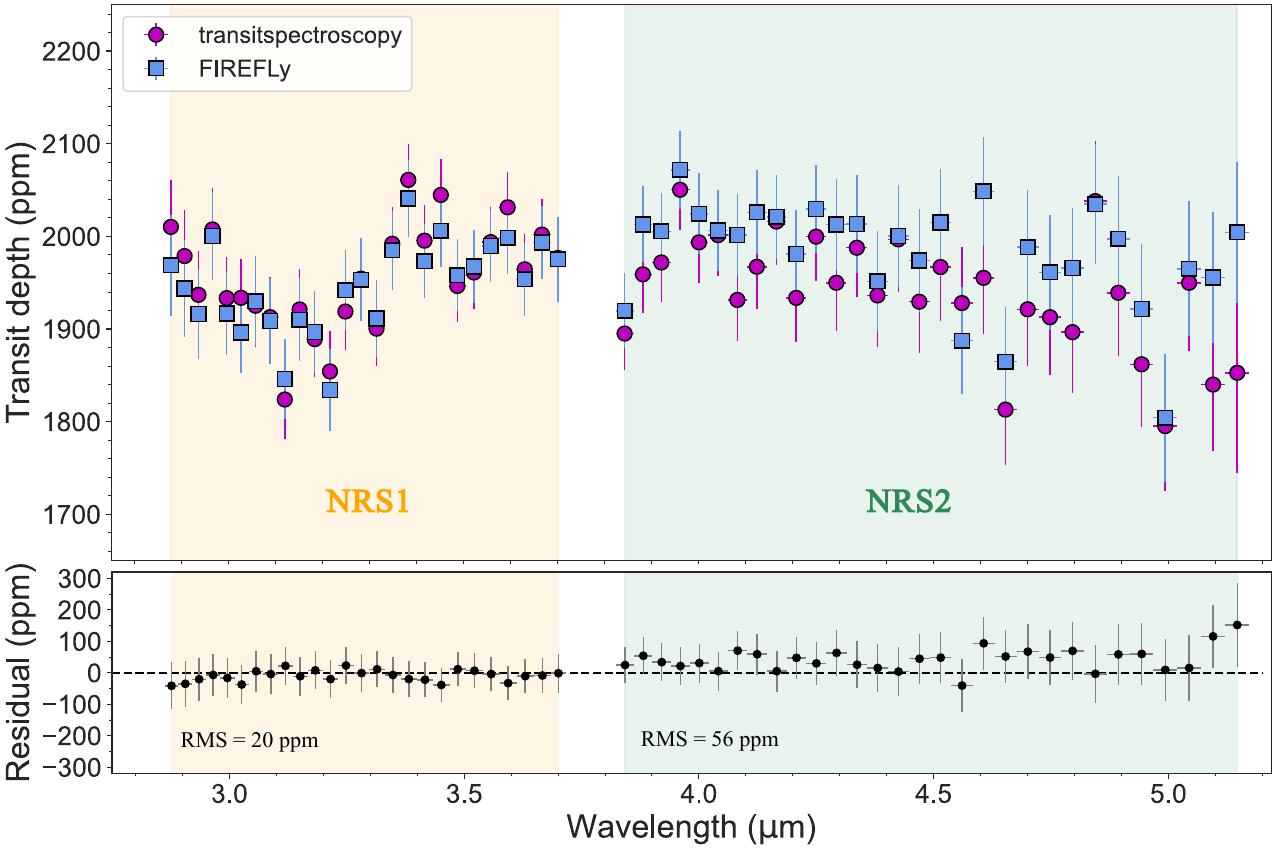}
    \caption{Transmission spectra of L98-59\,d obtained using \texttt{ transitspectroscopy} (pink) and  \texttt{FIREFLy}  (blue) pipelines. The spectral light curves from the two indepedent reductions were fitted using \texttt{ juliet} with a similar parametrization at pixel resolution and then binned to R $\sim$ 100. The residuals between the two reductions show a global offset between the two reductions for the NRS2 detector.}
    \label{fig:transit_spectra}
\end{figure*}

We independently fitted white and spectral light curves from each detector using the \texttt{juliet} Python package \citep{Espinoza_2019}, employing nested sampling via \texttt{dynesty} \citep{Speagle_2020}. We restricted the time series observations to +1.0 hours from the start, as the transit begins in the first half-hour (see top panel Figure\,\ref{fig:wlc}). The period ($P = 7.4507245$ days), eccentricity ($e = 0$), and argument of periastron ($\omega =90$), are fixed based on values from \citep{Luque_2022, Demangeon_2021}. We fitted the planet-to-star radius ratio, mid-transit time, impact parameter, and semi-major axis ratio, including limb-darkening coefficients determined with a square-root law and a uniform prior. Additionally, we included a mean-out-of-transit offset ($mflux$) and a jitter parameter($sigma\_w$) for white noise. To account for correlated noise we used Gaussian Processes (GP) using the \texttt{george} package with a Matèrn 3/2 kernel \citep{Ambikasaran_2015, Foreman_Mackey_2017}. This method captures complex systematics and has been validated in ERS WASP-39b studies \citep{Feinstein_2023, Alderson_2023}. A log-uniform prior was set for the GP amplitude between 0.001 and 1000 ppm, and an exponential prior was set for the GP length-scale, using time and the full-width half maximum (FWHM) as GP regressors. The white light curve fitting results are shown in Figure\,\ref{fig:wlc}, and the detailed best-fit parameters and confidence intervals are provided in Appendix Table\,\ref{tab:wlc_best_fit}. We determined the optimal systematics detrending model based on Bayesian log-evidence, finding that the GP model with both time and FWHM as regressors offers the best fit. We evaluated several detrending approaches, including linear models with time and FWHM, and a GP with time only using the Matèrn 3/2 kernel via the \texttt{celerite} package \citep{Foreman_Mackey_2017}. For NRS1, the log-evidence values were 6700 for the transit-only model, 6778 for the linear time model, 6780 for the linear time plus FWHM model, 6782 for the GP with time only, and 6784 for the GP with both time and FWHM. Similarly, for NRS2, the GP model including both time and FWHM achieved the highest log-evidence value of 6405, compared to 6302 for the transit-only model, 6212 for the linear time model, 6219 for the linear time plus FWHM model, and 6282 for the GP with time only.

For the spectral light curves, we used a similar setup but fixed the mid-transit time, impact parameter, and semi-major axis ratio based on the best-fit results from the white light curves. The fitted values are $T_0$ (BJD$_{\rm TDB}$) $= 2460121.112518755 \pm 7 \times 10^{-5}$, $b = 0.925\pm 0.004$, and ${a}/{R_{\star}} = 37.11 \pm 0.55$. Limb-darkening coefficients were determined using a truncated normal distribution centered on values estimated by the \texttt{ exotic} \citep{grant_2022_7437681}  package and PHOENIX models \citep{husser2013new}. Given the high impact parameter, constraining these coefficients was challenging, as discussed in Appendix\,\ref{Appendix:ldc}. The transmission spectrum was obtained by binning pixel-level transit depths to a resolution of R$\sim$100, resulting in 56 data points. Uncertainties in the final spectrum were calculated by determining the weighted mean and variance for each bin, with errors derived from the weighted contributions of the high-resolution spectra's errors and their associated variance. This spectrum can be found in Appendix\,\ref{tab:transit_depths}.

\subsection{Data reduction with  \texttt{FIREFLy} } \label{data_redu_firefly}
We reduced the L98-59 d data using the Fast InfraRed Exoplanet Fitting Lyghtcurve (\texttt{FIREFLy}; \citealt{Rustamkulov_2022,Rustamkulov_2023})
reduction suite. \texttt{FIREFLy} starts with a customised reduction using the STScI pipeline and the uncalibrated images and includes 1/$f$ destriping at the group level before the ramp is fit. We applied a custom superbias step, where the STScI pipeline superbias was scaled separately for NRS1 and NRS2 to match the trace-masked count level median value of each integration. This procedure was introduced in \cite{Moran_2023} to help reduce offsets between the NRS1 and NRS2 detectors.  The jump-step and dark-current stages of the STScI pipeline are skipped. We then use the custom-run 2D images after the gain scale step, and perform customised cleaning of bad pixels, cosmic rays and hot pixels.

We fit the transit light curves using a quadratic limb-darkening model ($q_1$, $q_2$; \citealt{kipping2013uninfomativePriorsQuad}, along with a polynomial function of time to remove overall baseline trends. In addition, we tested detrending the lightcurves with the X and Y positions on the detector and the scaling factor used in the superbias step. We found the NRS1 detector to have a step trend which evolved over the first 0.04 days requiring a polynomial in time with orders 1, 2 , 5 and 6. The higher-order trends modelled the trends seen at the beginning of the observation well.  Conversely, NRS2 only required a second-order polynomial though a trend in Y-position was found to be needed. In addition a trend for both NRS1 and NRS2 with the superbias scaling factor was found to be needed.  We first fit for both $q_1$ and $q_2$, but found $q_2$ to be weakly constrained and consistent with 0, so $q_2$ was fixed to zero throughout the rest of the analysis.  ${a}/{R_{\star}}$, $T_0$, and the impact parameter $b$ were fit for both NRS1 and NRS2, then fixed to the weighted-average value of both detectors.  The spectroscopic fits followed the same procedure.  However, $q_1$ was found not to have a strong wavelength dependence when fit spectroscopically and was weakly constrained, so it was fixed to the weighted average value for all wavelength channels ($q_1$=0.117$\pm$0.017). The lightcurves were binned to produce a final transmission spectra with a resolution near $R=$60 with a total of 38 points.

\subsection{Data reductions comparison}
We fit the pixel-light curves from \texttt{ transitspectroscopy} and  \texttt{FIREFLy} reductions with the parametrization described in Section\,\ref{sec:transitspectroscopy}. The results are in Figure\,\ref{fig:transit_spectra}. While both spectra exhibit similar shapes, a systematic offset is observed for NRS2 detector transit depths between the two reductions, with  \texttt{FIREFLy}'s transmission spectrum exhibiting higher transit depth values than \texttt{ transitspectroscopy}. Specifically, there is a higher median of residuals for NRS2 ($+46$ppm) compared to NRS1 ($-8$ppm), potentially attributable to a different background treatment for NRS2 detector. The root mean square of the residuals is found to be 56 ppm for NRS2 and 20 ppm for NRS1. Similarly, the light curve fitting setup described in Section\,\ref{data_redu_firefly} is applied to \texttt{ transitspectroscopy}'s light curves (see Appendix\,\ref{fig:transit_spectra_r60}). 

\begin{figure*}[htpb]
    \centering
  \includegraphics[width =\textwidth]{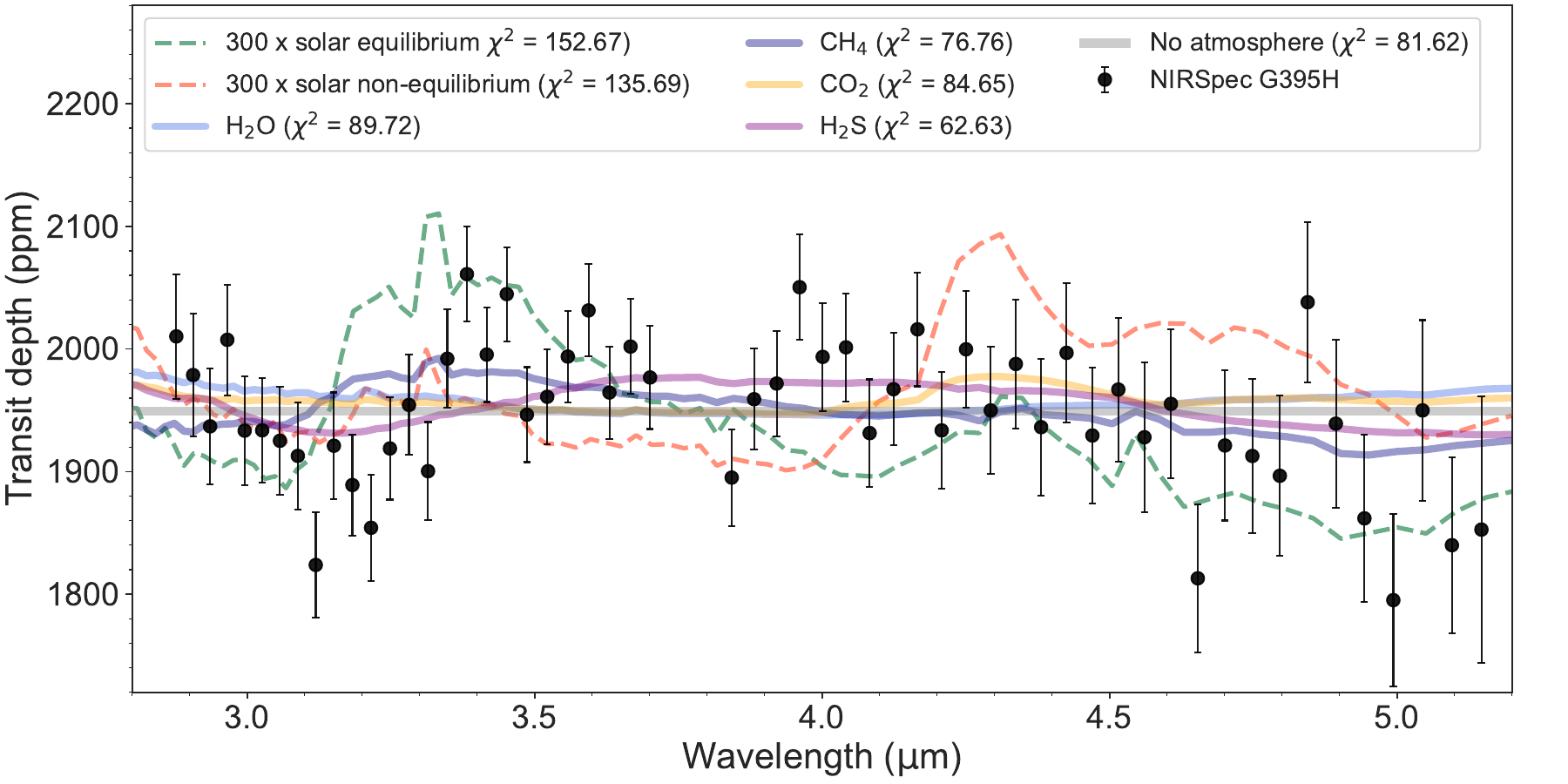}
    \caption{Transmission spectrum of L98-59\,d obtained using \texttt{ transitspectroscopy} and binned to R $\sim$ 100 (black points) compared to forward models from \texttt{ Exo-REM} (coloured dashed and filled lines). The pure H$_2$S model provides the best-fit according to $\chi^2$ statistics. No offset is added between NRS1 and NRS2. }
    \label{fig:exorem}
\end{figure*}

\section{Interpretation} \label{sec:models}
We interpret the results using forward and retrieval modelling focusing on the transmission spectrum obtained with \texttt{ transitspectroscopy} at R$\sim$100. 

\subsection{Evidence of an atmosphere from 1D forward modelling}
We used \texttt{Exo-REM} \citep{Baudino_2015, Charnay_2018, Blain_2021} to create a grid of transmission spectra for L98-59\,d. \texttt{Exo-REM} is a 1D self-consistent radiative-convective code for exoplanets and brown dwarfs' atmospheres. The atmosphere is modelled using 80 layers between 10$^{-3}$ and 10$^7$ Pa, with an Eddy Diffusion coefficient of 10$^{-8}$ cm$^2$/s$^{-1}$. We simulated a 300 $\times$ solar atmosphere, including 13 absorbing species in a hydrogen-rich atmosphere, considering both equilibrium and out-of-equilibrium chemistry. The metallicity is the factor by which all the elemental abundances except H are multiplied compared to their solar abundances \citep{Lodders_2010}. For out-of-equilibrium chemistry, \texttt{Exo-REM} incorporates disequilibrium processes for CH$_4$–CO, CO–CO$_2$, and N$_2$–NH$_3$–HCN reactions based on \citet{Zahnle_2014}. The quenching level is determined by comparing the reaction timescale to the mixing time H$^2$/K$_{zz}$, where H is the atmospheric scale height and K$_{zz}$ the Eddy diffusion coefficient. The abundance of the species is governed by the thermochemical equilibrium below the quenching level, where the temperature and pressure are high enough so that the kinetics dominates. The internal temperature was set to 70\,K, controlling the position of the convective layer.  We also tested a pure H$_2$O, CH$_4$, CO$_2$ and H$_2$S atmosphere.

Figure\,\ref{fig:exorem} shows the transmission spectrum obtained with \texttt{transitspectroscopy} and grid models from \texttt{Exo-REM}, along with corresponding $\chi^2$ statistics. The pure H$_2$S model provides the best fit to the spectrum with a chi-square difference of around 19 with the flat-line. This represents a statistically significant improvement in fit, exceeding 3$\sigma$ (dof\,=\,56). However, the observed variations in the transmission spectrum for a pure H$_2$S model, which show differences of approximately 40 ppm, exceed the simple scale height prediction of $\sim$5 ppm. This is estimated using a 416\,K equilibrium temperature, 34 g/cm$^3$ mean molecular weight. This discrepancy may arise from several factors including the modeling of the atmosphere with ExoREM, which simulates a pure H$_2$S atmosphere within a hydrogen-dominated framework, potentially resulting in an inflated effective scale height. Additionally, the use of cross-section opacities made for hydrogen dominated atmosphere and detailed atmospheric physics in our models may explain the larger signal compared to the simplified scale height calculation. Further investigation into the support for sulfur-bearing species opacities in the atmospheric model will be conducted using a Bayesian framework in Section\,\ref{sec:4.2}. Cloud-free models with 300 times solar abundance exhibit large CH$_4$ features at 3.3 $\upmu$m and CO/CO$_2$ features around 4.5$\upmu$m, which are not observed in the L98-59\,d transmission spectrum. Despite the pure H$_2$S model being a likely non-physical atmospheric model, it is the best-fit according to $\chi^2$ statistics. The modeled flat-line presented in Figure \,\ref{fig:exorem} is computed using the results of the retrieval analysis with \texttt{TauREx} for consistency (see Section \ref{sec:4.2}). Using a weighted mean of spectrum points yields a chi-squared of 80.47, instead of 81.62. This results might indicate an atmosphere with the presence of sulfur-bearing species or stellar contamination. 

\subsection{Atmospheric and stellar contamination retrieval analyses} \label{sec:4.2}

\subsubsection{Retrieval configurations}
We conduct an atmospheric and a stellar contamination retrievals of L98-59\,d's properties based on its observed spectra obtained through \texttt{ transitspectroscopy} using \texttt{ TauREx 3} \citep{Al_Refaie_2021}\footnote{\url{https://github.com/ucl-exoplanets/TauREx3_public}} and \texttt{ exoretrievals} \citep{Espinoza_2019}. \\

For the atmospheric retrieval analysis, we used the \texttt{Multinest} algorithm \citep{Feroz_2009, Buchner_2014} with 100 layers in the atmosphere, spanning from 10$^{-3}$ to 10$^{6}$ Pa. The model included an evidence tolerance of 0.5 and 1500 live points. Stellar parameters were fixed (radius: 0.303 R$_{\odot}$, T$_{\rm star}$: 3415 K, metallicity: -0.46). The planetary radius was fitted within the range of 0.79 to 2.37 R$_{\oplus}$ to explore the full parameter space, account for uncertainties including limb-darkening effects, and ensure robustness in the model. The temperature-pressure profile ranged from 200 to 600\,K. Clouds were modeled as grey clouds with a top pressure between 10$^{-3}$ and 10$^{6}$ Pa. The helium-to-hydrogen ratio was set to 0.17, and an offset between NRS1 and NRS2 transmission spectra was fitted uniformly between -100 and +100 ppm using the \texttt{ taurex-offset} plugin \footnote{\url{https://github.com/QuentChangeat/taurex-offset}}. \\

\textbf{Equilibrium chemistry}:
We used \texttt{ggchem} \footnote{\url{https://pypi.org/project/taurex-ggchem/}} \citep{Woitke_2018} for equilibrium chemistry, including molecular line lists and continuum from the ExoMol project \citep{Tennyson_2016, Chubb_2021}, HITEMP \citep{Tennyson_2018}, and HITRAN \citep{Rothman_1987, Rothman_2010}. The active molecules included are : H$_2$O \citep{Polyansky_2018}, CO \citep{Li_2015}, CO$_2$ \citep{Rothman_2010}, CH$_4$ \citep{Yurchenko_2017}, NH$_3$ \citep{Yurchenko_2011}, H$_2$S \citep{Azzam_2016}, HCN\citep{Barber_2014}, C$_2$H$_2$\citep{Chubb_2020}, C$_2$H$_4$\citep{Mant_2018}, SiO\citep{Yurchenko_2021}, SO$_2$ \citep{Underwood_2016}, TiO\citep{McKemmish_2019}, and VO\citep{McKemmish_2016}. Collision-induced absorption (CIA) and Rayleigh scattering were also considered. We fitted for C/O and S/O ratios (0.1 to 2) and atmospheric metallicity (1 to 1000 times solar). \\

\textbf{Free chemistry}:
The active molecules included are H$_2$O, CO, CO$_2$, CH$_4$, SiO, H$_2$S, HCN, NH$_3$, and SO$_2$. CIA and Rayleigh opacities were included. We fitted the abundance of each molecule ($\log(\text{X}_{\rm VMR})$) between 10$^{-12}$ and 0.3 (between -12 and -0.5 in log space), ensuring the total molecular abundances never exceed one. \\

\textbf{Free chemistry with N$_2$}:
The setup is similar to the free chemistry retrieval but included N$_2$ as an inactive gas. The N$_2$/H$_2$ ratio was fitted between 0.001 and 2. \\

\textbf{Flat-line}:
The flat-line is a retrieval with no opacity sources, fitting for planetary radius and temperature. This was used to assess the significance of the atmospheric detection using Bayesian evidence, evaluated using the Bayes's theorem. The Bayes factor, positively designed is computed between an atmospheric model and the flat-line and then converted in $\sigma$ significance using the formalism of \citet{Trotta_2008} and \citet{Benneke_2013}. This retrieval is performed for each atmospheric retrieval on the transmission spectrum with the retrieved offset applied and fixed.

\begin{figure*}
    \centering
  \includegraphics[width =17.3cm]{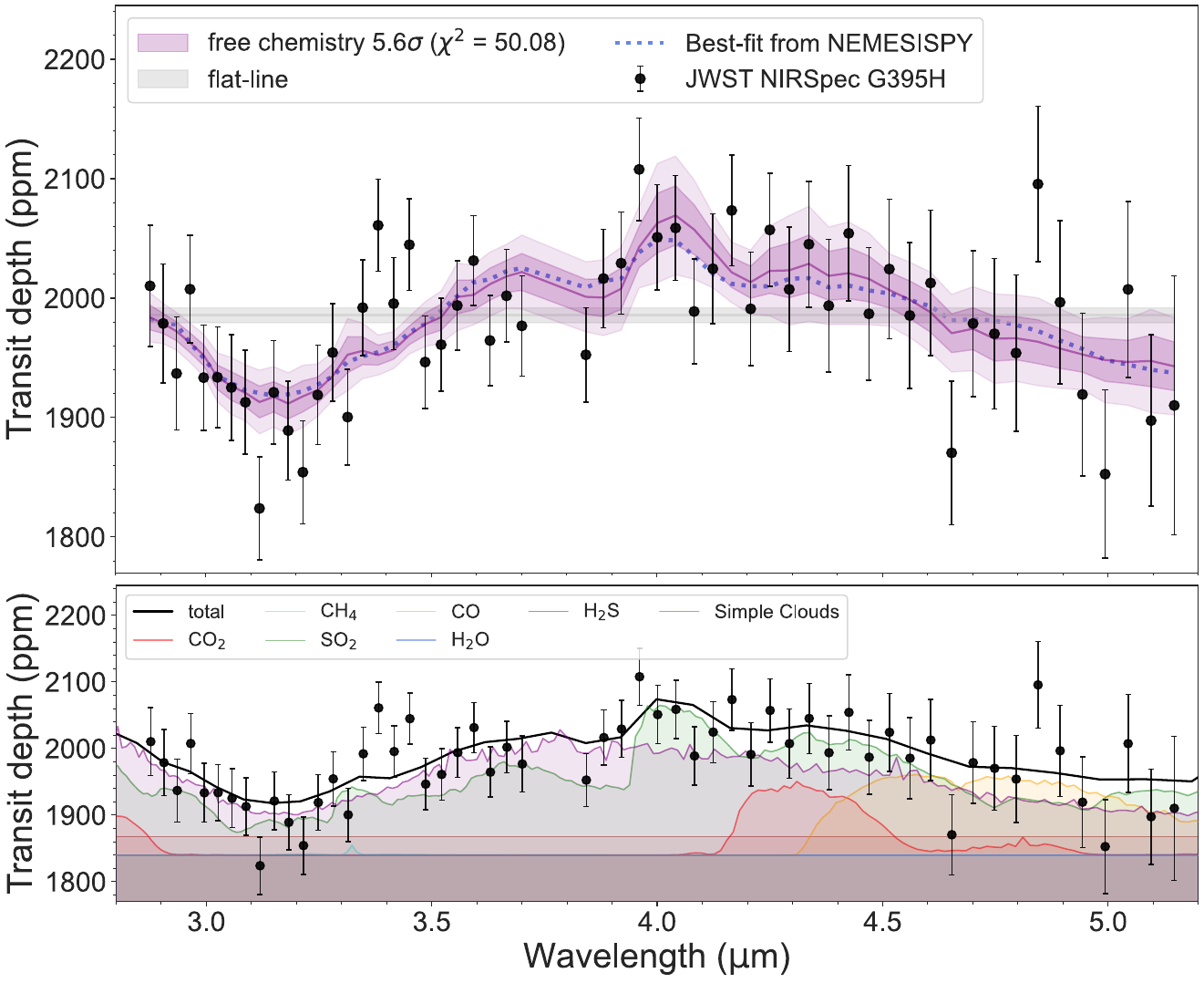}

 \includegraphics[width =17.3cm]{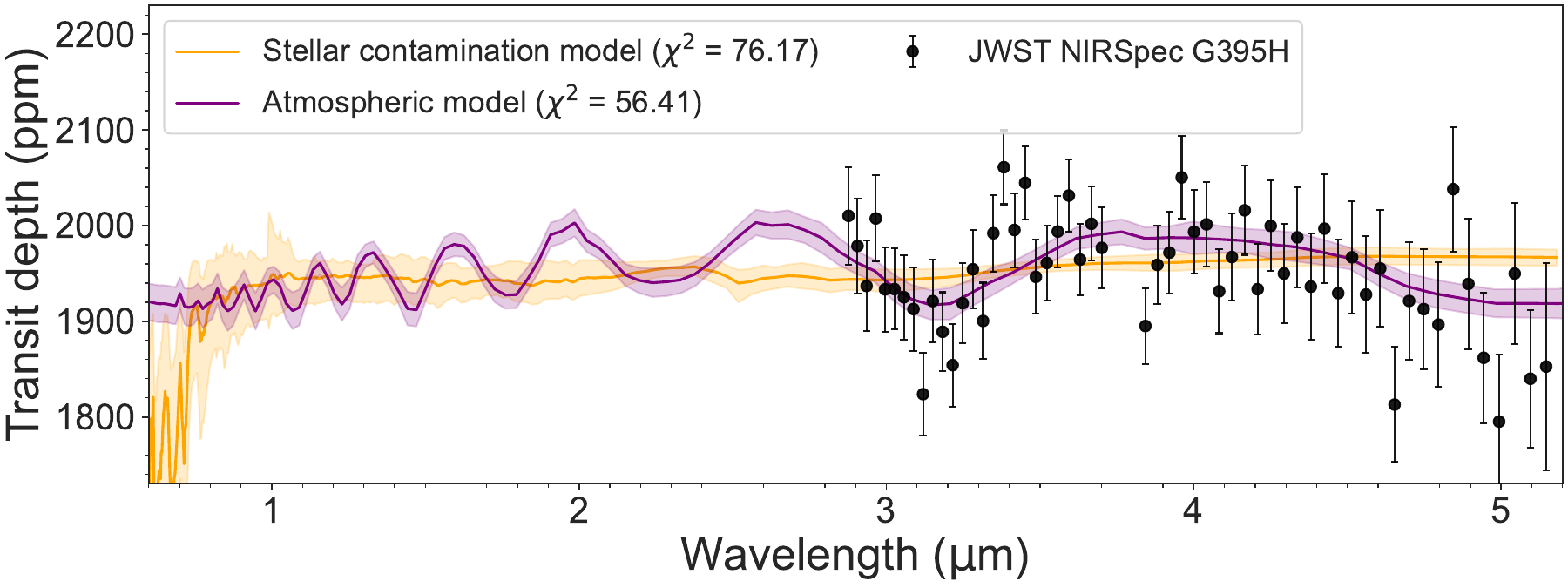}
    \caption{Top panel: Transmission spectrum of L98-59\,d obtained using \texttt{ transitspectroscopy} and binned to R $\sim$ 100 (black points) compared to a free chemistry retrieval model from \texttt{ TauREx} (purple) and \texttt{ NEMESISPY} (dashed blue) from Banerjee et al. 2024, hereafter Paper \RomanNumeralCaps{2}. Middle : Opacity contributions from the best-fit model. The model suggests a contribution from H$_2$S (purple) and SO$_2$ (green) opacities. Bottom : Transmission spectrum compared to stellar (yellow) and atmospheric (purple) retrievals models.}
    \label{fig:atmo_retrieval}
\end{figure*}

\begin{table*}[htpb]
  	\caption{Summary of Atmospheric retrieval results using \texttt{ TauREx} on L98-59\,d transmission spectrum obtained with \texttt{ transitspectroscopy}'s reduction at R$\sim$100. }         
  	\label{table:retrieval}     
  	\centering                                     
  	\begin{tabular}{ l |c c |c c |c c |c c}        
  	\hline\hline  & \multicolumn{2}{c}{equilibrium}
   &\multicolumn{2}{c}{free chemistry} &\multicolumn{2}{c}{free - with N$_2$} &\multicolumn{2}{c}{free - H$_2$S and SO$_2$ only}\\ 
    \hline
    Parameters & Prior &  Posterior &  Prior &  Posterior &  Prior &  Posterior &  Prior &  Posterior \\
    \hline 
    R$_{\rm P, ref}$ (R$_{\oplus}$)  & $\mathcal{U}(0.79, 2.37)$ &$1.45^{+0.01}_{-0.01}$ & $\mathcal{U}(0.79, 2.37)$ &  $1.43^{+0.01}_{-0.01}$ & $\mathcal{U}(0.79, 2.37)$ & $1.43^{+0.01}_{-0.01}$ & $\mathcal{U}(0.79, 2.37)$ & $1.43^{+0.01}_{-0.01}$  \\
    T$_{\rm P}$ (K) & $\mathcal{U}(200, 600)$ & $ 490^{+74}_{-134}$ & $\mathcal{U}(200, 600)$ &  $ 426^{+111}_{-105}$ & $\mathcal{U}(200, 600)$ &  $521^{+54}_{-86}$ & fixed  & $ 416$  \\
    log(P$_{\rm clouds}$) (Pa) & $\mathcal{U}(-3,6)$ & $4.18^{+1.16}_{-1.67}$ & $\mathcal{U}(-3,6)$ &  $5.16^{+0.54}_{-0.75}$& $\mathcal{U}(-3,6)$ & $5.17^{+0.52}_{-0.61}$  & $\mathcal{U}(-3, 6)$ &  $5.27^{+0.50}_{-0.61}$  \\
    mmw (amu) & - & $32.13^{+1.50}_{-8.31}$ & - &$9.18^{+2.51}_{-2.41}$  &- & $13.71^{+4.04}_{-3.46}$ & -& $9.30^{+2.46}_{-1.59}$ \\
     Offset (ppm)  & $\mathcal{U}(-100, 100)$ &$14.6^{+14.6}_{-14.1}$ & $\mathcal{U}(-100,100)$ & $37.6^{+15.0}_{-16.4}$ & $\mathcal{U}(-100, 100)$& $35.5^{+14.1}_{-16.0}$ & $\mathcal{U}(-100, 100)$ & $31.4^{+13.4}_{-13.6}$ \\ \hline
 
    Z &$\mathcal{U}(1, 1000)$ &$ 747^{+167}_{-239}$ & & & &\\
    C/O &$\mathcal{U}(0.1, 2)$ & $ 1.23^{+0.51}_{-0.58}$ &  & & & & \\
    S/O &$\mathcal{U}(0.1, 2)$ & $ 1.54^{+0.32}_{-0.52}$  &  & & & &\\

    \hline
    
    N$_2$/H$_{2}$ & &  & & & $\mathcal{U}(0.001, 2)$ & $ 0.64^{+0.62}_{-0.35}$  & &\\
    log(H$_{2}$O) & - & - & $\mathcal{U}(-12, 0.3)$ &  $<-5.0$  &$\mathcal{U}(-12, 0.3)$ &  $<-5.0$ & - & - \\
    log(CO) & - & -& $\mathcal{U}(-12, 0.3)$ &  $<-2.5$  & $\mathcal{U}(-12, 0.3)$ &  $<-2.5$  & - & - \\
    log(CO$_2$)& - & - & $\mathcal{U}(-12, 0.3)$ &  $<-2.5$ & $\mathcal{U}(-12, 0.3)$ &  $<-2.5$ & - & - \\
    log(CH$_4$)& - & - & $\mathcal{U}(-12, 0.3)$ &  $<-6.0$ & $\mathcal{U}(-12, 0.3)$ &  $<-6.0$   &- & - \\
    log(SiO) & - & -& $\mathcal{U}(-12, 0.3)$ &  $<-2.0 $ &$\mathcal{U}(-12, 0.3)$ & $<-1.0$  & - & - \\
    log(H$_2$S) & - & -& $\mathcal{U}(-12, 0.3)$ &  $-0.74^{+0.14}_{-0.49}$ &  $\mathcal{U}(-12, 0.3)$ &  $-1.31^{+0.48}_{-2.35}$ & $\mathcal{U}(-12, 1)$ &  $-0.66^{+0.13}_{-0.11}$   \\
    log(HCN) & - & -& $\mathcal{U}(-12, 0.3)$ &  $<-5.0$  & $\mathcal{U}(-12, 0.3)$ &  $<-5.0$  & - & -  \\
    log(NH$_3$) & - & -& $\mathcal{U}(-12, 0.3)$ &  $<-6.0$ & $\mathcal{U}(-12, 0.3)$ &  $<-6.0$ &- & -  \\
    log(SO$_2$) & - & -& $\mathcal{U}(-12, 0.3)$ &  $-5.64^{+4.41}_{-4.12}$ & $\mathcal{U}(-12, 0.3)$ &  $-3.80^{+2.25}_{-4.91}$ &  $\mathcal{U}(-12, 1)$ &  $<-2.0$ \\ \hline \hline

    Criteria & flat-line\footnote{The statistics on the flat-line model is computed for each spectrum with the fitted offset applied.} &  model & flat-line & model  & flat-line & model  &  flat-line & model  \\ \hline
   $\chi^2$ &  83.05 & 70.37 & 96.54 & 49.68 & 86.00 & 50.08 & 86.24 & 50.74 \\
    AIC &  91.05 & 84.38 & 104.54 & 75.68 & 94.00 & 78.08& 94.24 & 60.74 \\
    BIC &  99.15 & 98.55 & 112.64 & 102.01 &102.10 & 106.44 &102.34 & 70.87 \\
    log$\mathcal{Z}$& 458.00  & 460.31 & 451.18  & 464.99 &456.43& 464.92 &456.36 & 468.53 \\
    significance ($\sigma$)& - & 2.7 & - & 5.6 &-& 4.5 &-& 5.3  \\ \hline
  	\end{tabular}
\end{table*}

\subsubsection{Retrieval results : an atmosphere around L98-59d ?}
The retrieval results using \texttt{ transitspectroscopy} reduction are summarised in Table\,\ref{table:retrieval}. We detected an atmosphere around L98-59\,d at 5.6$\sigma$ with the free chemistry retrieval and at 2.7$\sigma$ with the equilibrium chemistry retrieval. Both retrievals report a high abundance of H$_2$S: $\log$(H$_2$S)=$-0.74^{+0.14}_{-0.49}$ (free chemistry) and $\log$(H$_2$S)=$-0.21^{+0.07}_{-0.09}$ (equilibrium), with Bayesian evidences $\log\mathcal{Z} = 464.99$ and $\log\mathcal{Z} = 460.31$, respectively. The flat-line fits yield $\log\mathcal{Z} = 451.19$ and $\log\mathcal{Z} = 458.00$. \\

The free chemistry retrieval best-fit model is presented in Figure\,\ref{fig:atmo_retrieval} and the corresponding posterior distributions are in Appendix\,\ref{fig:atmo_retrieval_free_post} while the equilibrium chemistry retrieval is presented in Appendix\,\ref{fig:atmo_retrieval_eq}. The results align with Paper \RomanNumeralCaps{2} using NEMESISPY\citep{Irwin_2008, yang_2023}. An offset of $37.6^{+15.0}_{-16.4}$ ppm was found between NRS2 and NRS1 with transit depths higher for NRS2. Both models suggest significant H$_2$S opacity, with CO$_2$ (equilibrium) and SO$_2$ (free chemistry) as additional opacity contributers. The free chemistry model fits significantly better than the equilibrium model by 3.8$\sigma$ suggesting that L98-59\,d's atmosphere is not at equilibrium. The SO$_2$ VMR in the free retrieval shows a bi-modal distribution, with a constraint either above $-1.0$ or an upper limit below $-1.5$. However, we can put upper limits on every other molecules of the fit. In particular, H$_2$O's VMR is constrained below $10^{-5}$ suggesting a water-poor atmosphere for this planet that has density compatible with a water-rich composition \citep{Luque_2022}. Excluding both H$_2$S and SO$_2$ from the free chemistry retrieval indicates that the model with these molecules is favored at 4.5$\sigma$ using model comparison with the Bayes factor.

Adding N$_2$ as an inactive gas in the free chemistry retrieval did not significantly alter results, favoring the atmospheric model over a flat-line at 4.5$\sigma$. The only strongly constrained parameters are the planetary radius at 10 bars, the log VMR of H$_2$S, toward the edge of the prior (0.3), and the top pressure of the grey clouds. The log VMR of SO$_2$ is tentatively constrained. For this reason, we run a fit with H$_2$S  and SO$_2$ as the only active molecular opacity sources, increasing the prior range to 1, and fit for their molecular abundances while fixing the temperature to the equilibrium temperature at 416K. This atmospheric model is preferred at 5.3$\sigma$ to a flat-line. 

\subsubsection{Stellar contamination : is there an atmosphere or stellar inhomogeneities ?}
One of the early results from observations of small planets with JWST is how important it is to consider heterogeneities on the surface of the host star when looking for atmospheric signals in transmission. Multiple small planet observations have shown contamination from the presence of hot and cold spots on the photosphere of the host star during transit, which imprint false slopes and can potentially even inject false molecular features into the transmission spectrum \citep[see e.g.][]{Moran_2023, Lim_2023}. This is especially important for planets orbiting small M stars like L98-59, which are known to show a high level of magnetic activity and associated surface features \citep{Rackham_2018}.

To test if the spectrum could be explained by stellar contamination caused by inhomogeneities (hot and/or cold spots) on the stellar surface, we carry out retrievals using exoretrievals \citep{Espinoza_2019}. Based on the framework presented in \citet{Rackham_2018}, this allows us to model the spectrum created by occulted and/or unocculted spots on the stellar surface during the transit, by replacing a portion of the surface with a stellar model of another temperature. Since we do not see any evidence of spot crossing events in the light curves, we only consider unocculted spots in our retrieval. We consider the cases of only hot spots, only cold spots, or both being present on the stellar surface, and allow these spot temperatures to vary over a large parameter space, from 1800 K to 3250 K and from 3550 K to 5000 K, respectively, for a photospheric temperature of $3415 \pm 135$. We also place a wide prior on the spot covering fraction, allowing the spots to together cover up to the entirety of the unocculted photosphere. We also tested using both default PHOENIX models and BT-SETTL models considering the relatively low photospheric temperature of L98-59, but saw no appreciable difference in the results. Even with this very large explored parameter space, none of the stellar contamination models are able to explain the features seen in the spectrum, especially those at the longest wavelengths.
Figure\,\ref{fig:atmo_retrieval} bottom panel shows the best-fit model from the stellar contamination retrieval results. The corresponding posterior distributions are in Appendix\,\ref{fig:stellar_retrieval_post}. We also show the best-fit model from an atmospheric retrieval analysis with \texttt{ TauREx}. In these fits, no offset between NRS1 and NRS2 is applied. The H$_2$S/SO$_2$ rich atmosphere is favored compared to the stellar contamination suggesting the large feature between 3.3 to 4.8$\upmu$m, if confirmed by follow-up observations is originated from the atmosphere. 

\subsubsection{Comparison between data reductions}
To confirm the atmospheric detection, we applied the free chemistry retrieval setup to the \texttt{FIREFLy} transmission spectrum at R $\sim$ 100 and to the spectrum obtained at R $\sim$ 60 from an independent fitting method (see Figure\,\ref{fig:transit_spectra}). The results of these fits are shown in Appendix Figure\,\ref{fig:retrieval_comparison}. The atmospheric detection was less significant with \texttt{FIREFLy}, yielding 3.0$\sigma$ for R $\sim$ 100 and 2.6$\sigma$ for R $\sim$ 60 spectra.

The discrepancy at R $\sim$ 100 primarily arises from differences at longer wavelengths in the NRS2 detectors. \texttt{transitspectroscopy} data showed a decrease in transit depth between 4.5 and 5$\upmu$m, fitted with H$_2$S opacity (see the middle panel in Figure\,\ref{fig:atmo_retrieval}) while \texttt{FIREFLy}'s transmission spectrum remains flat across the entire wavelength range of the NRS2 detector. However, sulfur-bearing species consistently appeared as the primary source of opacity in all retrievals. The NRS1 and NRS2 offset varied by reduction method: +37 ppm for \texttt{transitspectroscopy} and -14 ppm for \texttt{FIREFLy}, both consistent with zero at 1$\sigma$. Removing this offset is unlikely to significantly change the \texttt{FIREFLy} result given the small offset within 1$\sigma$ uncertainty.

The R $\sim$ 60 spectrum from \texttt{FIREFLy}  with an independent light curve fitting showed a tentative atmospheric detection (2.6$\sigma$). This finding raises concerns, suggesting that the detection made with other binning and light curve fitting may be attributed to random noise. Specifically, the discrepancy observed at shorter wavelengths, between 3 and 3.3$\upmu$m, between the two reductions could account for this result (see Figure\,\ref{fig:transit_spectra_r60}). The variability observed in \texttt{FIREFLy}'s transmission spectrum in this wavelength range resulted in a seemingly flatter spectrum and a statistically non-detection, whereas \texttt{transitspectroscopy}'s data indicated a pronounced decrease in transit depths, well-fitted by H$_2$S's opacity (see the middle panel in Figure\,\ref{fig:atmo_retrieval}). We note that employing the same light curve fitting and binning, the decrease is observed in \texttt{FIREFLy}'s reduction as well (see Figure\,\ref{fig:transit_spectra}). \\

In Appendix\,\ref{Appendix:retrieval}, we assess the impact of introducing an offset between NRS1 and NRS2 transit depths on atmospheric detection for L98-59\,d (see Appendix \,\ref{Appendix:retrieval:offset}) and evaluate the influence of spectral resolution (see Appendix \,\ref{Appendix:retrieval:resolution}).

\section{Conclusions} \label{sec:conclusions}
We presented one transit observation of the Super-Earth L98-59\,d with the JWST NIRSpec G395H mode. Our study highlights several challenges encountered in the data reduction process when analyzing transmission spectra of small planets. First, the limb-darkening coefficients pose a challenge for grazing planets due to their impact on the light curve fitting process. Even though our two data reductions agree when using the same limb-darkening coefficients, remaining discrepancies at specific wavelength led to different interpretation in the retrieval analysis.

From this single transit, our analysis suggests no stellar contamination but hints at potential atmospheric detection, though significance varies with data reduction methods. In the best case scenario, we detected an atmosphere with a 5.6$\sigma$ significance. An independent analysis led to a 2.6$\sigma$ tentative detection. Atmospheric retrieval models suggest the presence of sulfur-bearing species with hydrogen and helium as background gases, although the inferred high abundance levels are not yet well understood. The discovery of sulfur dioxide in the atmospheres of hot Jupiters \citep{Rustamkulov_2023, Alderson_2023, Powell_2024} and Neptune-like planets\citep{Dyrek_2023, benneke2024jwst, holmberg2024possible} are key-result in early JWST analysis being evidence of photo-chemistry. Sulfur species and clouds are expected in giant planets and brown-dwarf atmosphere \citep{Morley_2012, Tsai_2023, Polman_2023}. \citet{Crossfield_2023} showed that the ability to detect SO$_2$ in exoplanet atmospheres provides a crucial test for different planet formation models, revealing that sulfur's volatility and abundance can distinguish between planetesimal and pebble-accretion scenarios. Photochemistry could also produce H$_2$S and SO$_2$ in terrestrial atmosphere as it was predicted by \citet{Hu_2013}. The presence of sulfur-bearing species in rocky planets could also be explained by out-gassing, volcanism, interaction between the atmosphere and the rocky surface. The recent study of \citet{Janssen_2023} investigated sulfur's presence and detectability in rocky exoplanet atmospheres using thermochemical equilibrium models at the crust–atmosphere interface, considering surface temperatures of 500–5000 K and pressures of 1–100 bar, with various element abundances based on common rock compositions. They showed that at temperatures between 1000 and 2000\,K, gaseous sulfur concentrations can reach up to 25$\%$. They conclude that the most abundant sulfur molecules are SO, SO$_2$, H$_2$S, and S$_2$ with potentially detectable features in transmission spectra around 4$\upmu$m, 7–8$\upmu$m, and beyond 15$\upmu$m.  

The difference between the data reductions and the uncertainties on the data points are magnified when dealing with observations from a single visit. \citep{May_2023} highlighted for GJ\,1132\,b the importance of multiple visits when claiming atmospheric detection for rocky planets. Our transmission spectrum might have random noise fluctuations that suggested the detection of an atmosphere which emphasizes the need for multiple transits observations. This result will have to be compared to the second G395H visit from Cycle 2 GO program 4098 (PI: Benneke) but also to the NIRISS SOSS transit from Cycle 1 GTO program 1201 (PI: Lafrenière). 

Only hints of atmospheric detections have been found so far for rocky planets with radii below 1.6 R$_{\oplus}$ around M-dwarfs. Notably, planets from the GO program 1981 (PIs: Stevenson and Lustig-Yaeger), in order of increasing radius, include GJ\,341\,b with a radius of 0.92 R$_{\oplus}$ \citep{Kirk_2024}, LHS\,475\,b with 0.99 R$_{\oplus}$ \citep{Lustig_Yaeger_2023}, GJ\,1132\,b with 1.1 R$_{\oplus}$ \citep{May_2023}, and GJ\,486\,b with 1.3 R$_{\oplus}$ \citep{Moran_2023}, yet they have not provided compelling evidence of an atmosphere. However, similar JWST observations on planets with radii above 1.6R$_{\oplus}$, such as K2-18\,b (2.61 R$_{\oplus}$) \citep{madhusudhan2023carbonbearing} and TOI-270\,d (2.31R$_{\oplus}$), 55\,Cancri\,e (1.97 R$_{\oplus}$)\citep{Hu_2024} have provided strong molecular detections. If confirmed, the detection of sulfur-bearing species in an hydrogen-dominated atmosphere around L98-59\,d, a planet with a radius of 1.58 Earth radii, would be a significant result, as it lies right at the cutoff predicted by \citet{Rogers_2015, Rogers_2021} for planets to have retained their primary hydrogen-helium atmosphere. Planets with a radius above 1.6R$_{\oplus}$ have a density too low to be compatible with a silicate and iron-only core composition. A second study Paper \RomanNumeralCaps{2}, from the NIRSpec GTO collaboration will explore in more details the possible atmospheric scenarios for L98-59\,d.

\begin{acknowledgments}
N.H.A. acknowledges support by the National Science Foundation Graduate Research Fellowship under Grant No. DGE1746891.

The JWST data presented in this paper were obtained from the Mikulski Archive for Space Telescopes (MAST) at the Space Telescope Science Institute. The specific observations analyzed can be accessed via  \dataset[10.17909/nrxs-cx46]{http://dx.doi.org/10.17909/nrxs-cx46}.
\end{acknowledgments}

\facilities{JWST NIRSPEC G395H}
\bibliography{l9859d}{}
\bibliographystyle{aasjournal}

\appendix  
\renewcommand\thefigure{\thesection.\arabic{figure}}   
\section{Data reduction :  White light curve best-fit results}

\begin{table}[h!]
\centering
\caption{L98-59\,d White Light Curve fitting Results for the \texttt{transitspectroscopy} reduction using \texttt{juliet}.}
\footnotesize

\begin{tabular}{l|c|ccc|ccc}
\hline
\textbf{Parameter} & \textbf{Prior} & \multicolumn{3}{c|}{\textbf{NRS1}} & \multicolumn{3}{c}{\textbf{NRS2}} \\
\cline{3-8}
& & \textbf{Median} & \textbf{68\% CI Upper} & \textbf{68\% CI Lower} & \textbf{Median} & \textbf{68\% CI Upper} & \textbf{68\% CI Lower} \\
\hline
$T_0$ (BJD$_{\rm TDB}$) & $\mathcal{N}(2460121.1107, 0.2)$ & 2460121.1125 & +0.000076 & -0.000073 & 2460121.1125 & +0.000162 & -0.000168 \\
$a/R_{\star}$ & $\mathcal{N}(37.1, 0.5)$ & 37.098 & +0.449 & -0.450 & 37.117 & +0.101 & -0.100 \\
$b$ & $\mathcal{N}(0.92, 0.5)$ & 0.9230 & +0.0032 & -0.0034 & 0.9291 & +0.0078 & -0.0079 \\
$q1$ & $\mathcal{U}(0.0, 1.0)$ & 0.3722 & +0.274 & -0.162 & 0.4672 & +0.347 & -0.315 \\
$q2$ & $\mathcal{U}(0.0, 1.0)$ & 0.6287 & +0.259 & -0.356 & 0.5869 & +0.284 & -0.369 \\
$R_{\rm P}/R_{\star}$ & $\mathcal{U}(0.0, 0.2)$ & 0.0468 & +0.0012 & -0.0010 & 0.0385 & +0.0098 & -0.0110 \\
$mflux$ & $\mathcal{N}(0.0, 0.1)$ & -0.000025 & +0.000117 & -0.000120 & 0.000178 & +0.000477 & -0.000432 \\
$sigma\_w$ & $\text{log}\,\mathcal{U}(10^{-5}, 1000.0)$ & 142.50 & +4.48 & -4.44 & 258.14 & +8.11 & -7.52 \\
$GP\_sigma$ & $\text{log}\,\mathcal{U}(0.001, 1000.0)$ & 106.45 & +64.35 & -31.47 & 655.99 & +205.82 & -182.66 \\
$GP\_malpha0$ & $\text{Exp}(1.0)$ & 0.305 & +0.441 & -0.207 & 3.751 & +2.073 & -1.396 \\
$GP\_malpha1$ & $\text{Exp}(1.0)$ & 0.0202 & +0.0224 & -0.0124 & 0.0025 & +0.0030 & -0.0013 \\
\hline
\end{tabular}
\begin{tablenotes}
\footnotesize
\item Description of parameters:Mid-transit time (BJD$_{\rm TDB}$). Ratio of the semi-major axis to the stellar radius. Planet to star radius ratio. First limb-darkening coefficient of the squareroot law with the \citet{kipping2013uninfomativePriorsQuad} parametrization. Second limb-darkening coefficient. Out-of-transit flux offset. White noise jitter parameter. GP amplitude. GP length scales.
\end{tablenotes}
\label{tab:wlc_best_fit}
\end{table}

\section{Data reduction: treatment of the limb-darkening coefficients}\label{Appendix:ldc}

L98-59\,d had a high impact parameter $b>0.9$, which makes it challenging to constrain the limb-darkening coefficient in the light curve fitting. The choice of the limb-darkening coefficient treatment impact the extracted transit spectrum particularly the absolute transit depth value.  First, we ran tests: one involving fitting the limb-darkening coefficients uniformly between 0 and 1 using a quadratic law, and another using a square-root law. For this test, we fit the light curves from the \texttt{ transitspectroscopy}'s reduction at pixel resolution and bin the spectrum after. Results of this test are in Appendix Figure\,\ref{fig:quad_sq}. We note that there is a 400ppm offset between the two resulting transit spectra. The fit involving the quadratic law has a higher mean transit depth and lower precision. The shape of the spectrum is the same but the transit depth values and precision are sensitive to the choice of the limb-darkening treatment. 

Additionally, we perform a fit with fixed limb-darkening coefficients ($q_1=0.13$ and $q_2=0.0$), derived from the combined best-fit of NRS1 and NRS2 white light curves, with $q_2=0.0$ being unconstrained. This parametrization is applied to pixel-level \texttt{ transitspectroscopy} light curves, while  \texttt{FIREFLy}  light curves are binned to a R$\sim$ 60 and fitted with these fixed coefficients and an independent detrending linear model. This fitting is described in Section\,\ref{data_redu_firefly}. The results of this additionnal test is in Appendix\,\ref{fig:transit_spectra_r60}.

\begin{figure*}[htpb]
    \centering
  \includegraphics[width=\textwidth]{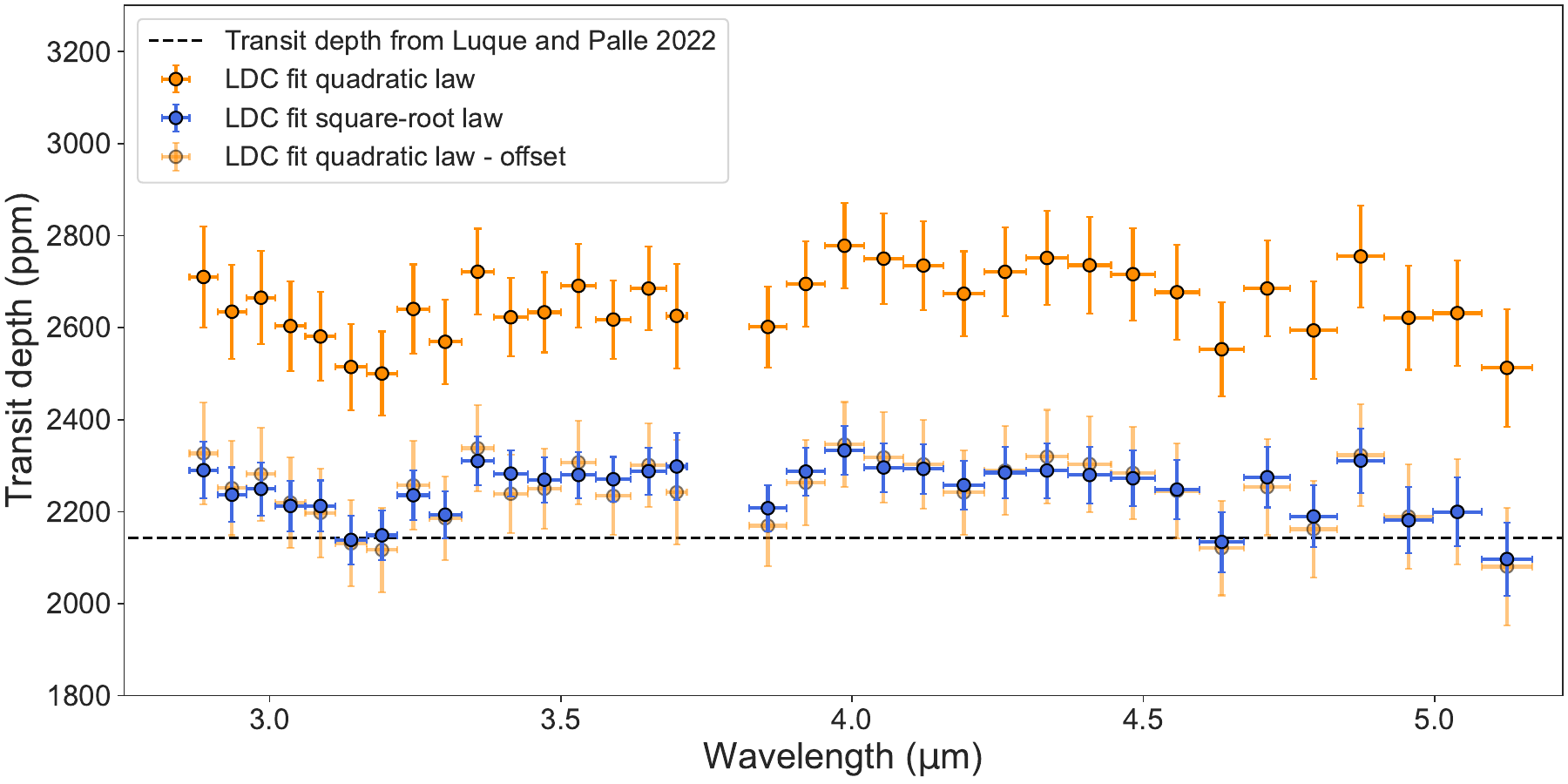}

  \includegraphics[width=\textwidth]{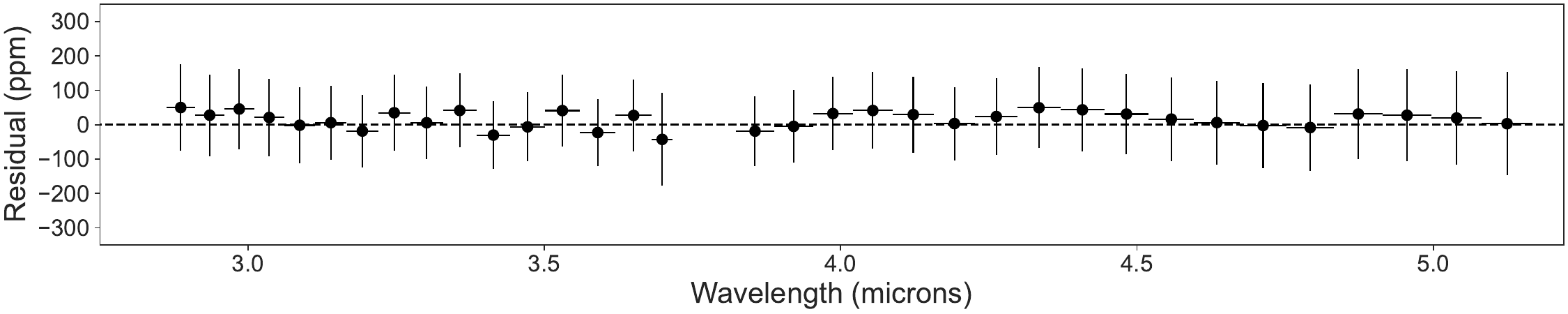}
    \caption{Top: Transmission spectra of L98-59\,d obtained using the reductions from \texttt{ transitspectroscopy} and binned to a resolution of R $\sim$ 60. The two spectra are derived by fitting the limb-darkening coefficients uniformly between 0 and 1 using a quadratic law (orange) and employing a square-root law (blue). The light orange points are obtained by applying an offset corresponding to the median differences between the two spectra. Bottom: Residuals between the two transmission spectra after subtraction of the median transit depth. Each detector is treated independently. The shape of the two spectra is similar but there is a 400ppm offset between the two reductions. }
    \label{fig:quad_sq}
\end{figure*}

\begin{figure*}[htpb]
    \centering
  \includegraphics[width=\textwidth]{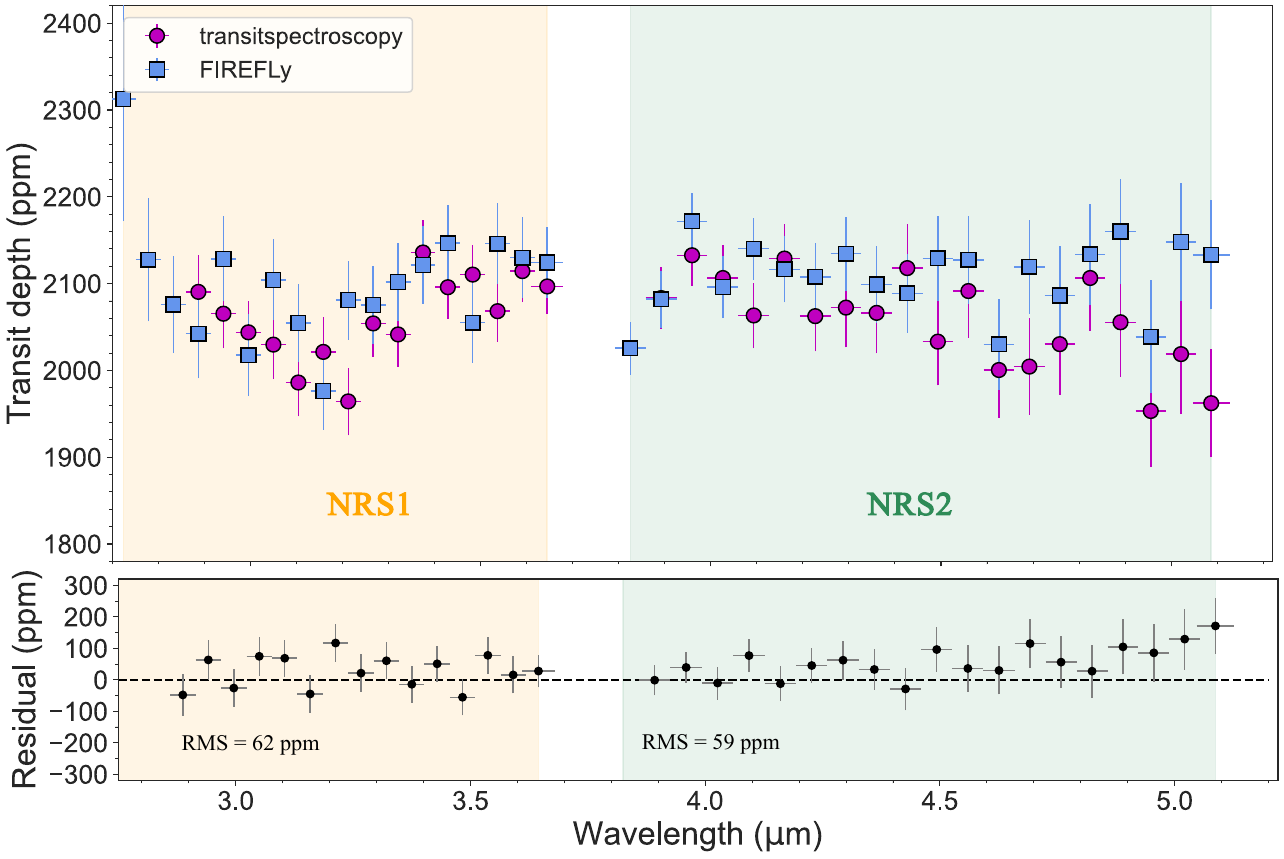}

    \caption{Transmission spectra of L98-59\,d obtained using \texttt{ transitspectroscopy} (pink) and  \texttt{FIREFLy}  (blue) pipelines. The spectral light curves from the two independent reductions were fitted independently. \texttt{ transitspectroscopy} are fitted at pixel-level resolution using Gaussian Processes to model the noise and then binned to the native resolution of the  \texttt{FIREFLy}'s reduction to approximately a R $\sim$ 60.  \texttt{FIREFLy}  light curves are binned to R $\sim$ 60 and then fitted using  a linear de-trending model. For both fitting the coefficients of the limb-darkening coefficients are fixed to $q_1=0.13$ and $q_2=0.0$.  }
    \label{fig:transit_spectra_r60}
\end{figure*}

\newpage

\begin{table}[h!]
\centering
\caption{L98-59\,d extracted transit depths from \texttt{transitspectroscopy} at R $\sim$ 100}
\footnotesize
\begin{tabular}{|c|c|c|c|}
\hline
Wavelength ($\upmu$m) & Transit Depth & Transit Depth error & Bin size ($\upmu$m) \\ \hline
2.8768 & 0.002010 & 0.000051 & 0.01460 \\ \hline
2.9060 & 0.001979 & 0.000050 & 0.01460 \\ \hline
2.9356 & 0.001937 & 0.000047 & 0.01495 \\ \hline
2.9654 & 0.002007 & 0.000045 & 0.01495 \\ \hline
2.9956 & 0.001933 & 0.000044 & 0.01525 \\ \hline
3.0261 & 0.001934 & 0.000042 & 0.01525 \\ \hline
3.0570 & 0.001925 & 0.000044 & 0.01560 \\ \hline
3.0881 & 0.001913 & 0.000043 & 0.01555 \\ \hline
3.1193 & 0.001824 & 0.000043 & 0.01560 \\ \hline
3.1509 & 0.001921 & 0.000043 & 0.01595 \\ \hline
3.1830 & 0.001889 & 0.000041 & 0.01620 \\ \hline
3.2483 & 0.001919 & 0.000042 & 0.01655 \\ \hline
3.2814 & 0.001954 & 0.000041 & 0.01655 \\ \hline
3.3146 & 0.001900 & 0.000040 & 0.01660 \\ \hline
3.3481 & 0.001992 & 0.000040 & 0.01690 \\ \hline
3.3822 & 0.002061 & 0.000039 & 0.01720 \\ \hline
3.4166 & 0.001995 & 0.000039 & 0.01725 \\ \hline
3.4514 & 0.002045 & 0.000038 & 0.01755 \\ \hline
3.4865 & 0.001946 & 0.000039 & 0.01750 \\ \hline
3.5219 & 0.001961 & 0.000039 & 0.01790 \\ \hline
3.5576 & 0.001994 & 0.000037 & 0.01785 \\ \hline
3.5937 & 0.002031 & 0.000038 & 0.01820 \\ \hline
3.6301 & 0.001964 & 0.000038 & 0.01820 \\ \hline
3.6668 & 0.002002 & 0.000039 & 0.01850 \\ \hline
3.7008 & 0.001977 & 0.000042 & 0.01550 \\ \hline
3.8428 & 0.001895 & 0.000040 & 0.01945 \\ \hline
3.8817 & 0.001959 & 0.000041 & 0.01945 \\ \hline
3.9209 & 0.001972 & 0.000043 & 0.01980 \\ \hline
3.9608 & 0.002050 & 0.000043 & 0.02010 \\ \hline
4.0010 & 0.001994 & 0.000044 & 0.02010 \\ \hline
4.0415 & 0.002001 & 0.000044 & 0.02045 \\ \hline
4.0827 & 0.001931 & 0.000044 & 0.02070 \\ \hline
4.1242 & 0.001967 & 0.000046 & 0.02075 \\ \hline
4.1659 & 0.002016 & 0.000046 & 0.02100 \\ \hline
4.2080 & 0.001934 & 0.000048 & 0.02105 \\ \hline
4.2504 & 0.001999 & 0.000047 & 0.02135 \\ \hline
4.2934 & 0.001950 & 0.000052 & 0.02165 \\ \hline
4.3370 & 0.001988 & 0.000053 & 0.02200 \\ \hline
4.3810 & 0.001936 & 0.000056 & 0.02195 \\ \hline
4.4252 & 0.001997 & 0.000057 & 0.02230 \\ \hline
4.4701 & 0.001929 & 0.000056 & 0.02255 \\ \hline
4.5152 & 0.001967 & 0.000058 & 0.02260 \\ \hline
4.5606 & 0.001928 & 0.000061 & 0.02285 \\ \hline
4.6067 & 0.001955 & 0.000061 & 0.02320 \\ \hline
4.6534 & 0.001813 & 0.000060 & 0.02350 \\ \hline
4.7007 & 0.001921 & 0.000061 & 0.02380 \\ \hline
4.7483 & 0.001913 & 0.000063 & 0.02375 \\ \hline
4.7961 & 0.001897 & 0.000065 & 0.02410 \\ \hline
4.8445 & 0.002038 & 0.000065 & 0.02435 \\ \hline
4.8936 & 0.001939 & 0.000068 & 0.02470 \\ \hline
4.9433 & 0.001862 & 0.000068 & 0.02495 \\ \hline
4.9935 & 0.001795 & 0.000070 & 0.02530 \\ \hline
5.0443 & 0.001950 & 0.000074 & 0.02555 \\ \hline
5.0954 & 0.001840 & 0.000072 & 0.02550 \\ \hline
5.1467 & 0.001853 & 0.000108 & 0.02585 \\ \hline
\end{tabular}
\label{tab:transit_depths}
\end{table}

\section{Additional retrieval analysis and Figures}\label{Appendix:retrieval}
\subsection{Free chemistry retrieval analysis}\label{Appendix:retrieval:free}
\begin{figure*}
    \centering
  \includegraphics[width =18cm]{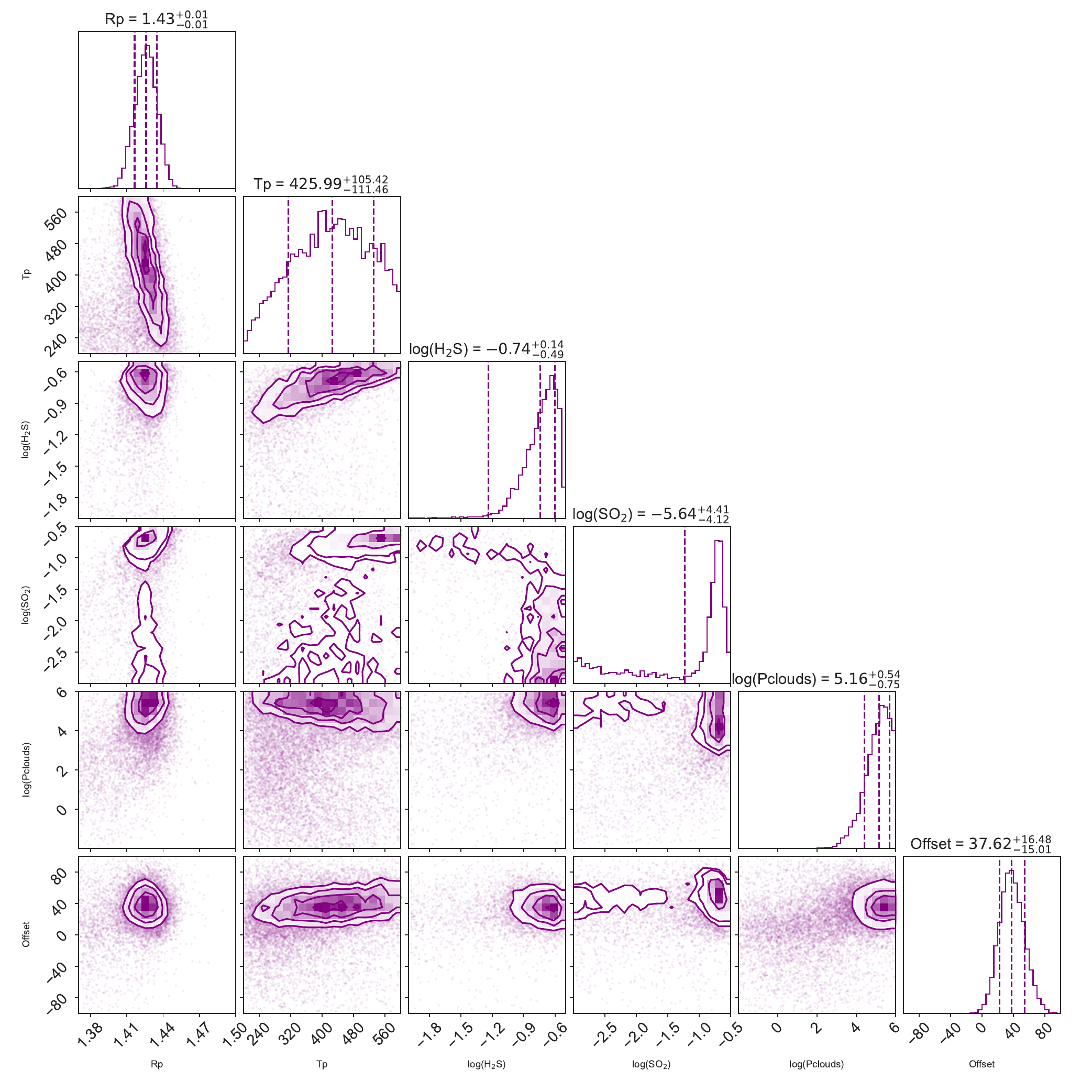}
    \caption{Posterior distributions from the free retrieval analysis. Not all parameters in the fit are plotted for clarity.  }
    \label{fig:atmo_retrieval_free_post}
\end{figure*}

\subsection{Equilibrium chemistry retrieval analysis}\label{Appendix:retrieval:eq}
\begin{figure*}
    \centering
  \includegraphics[width =18cm]{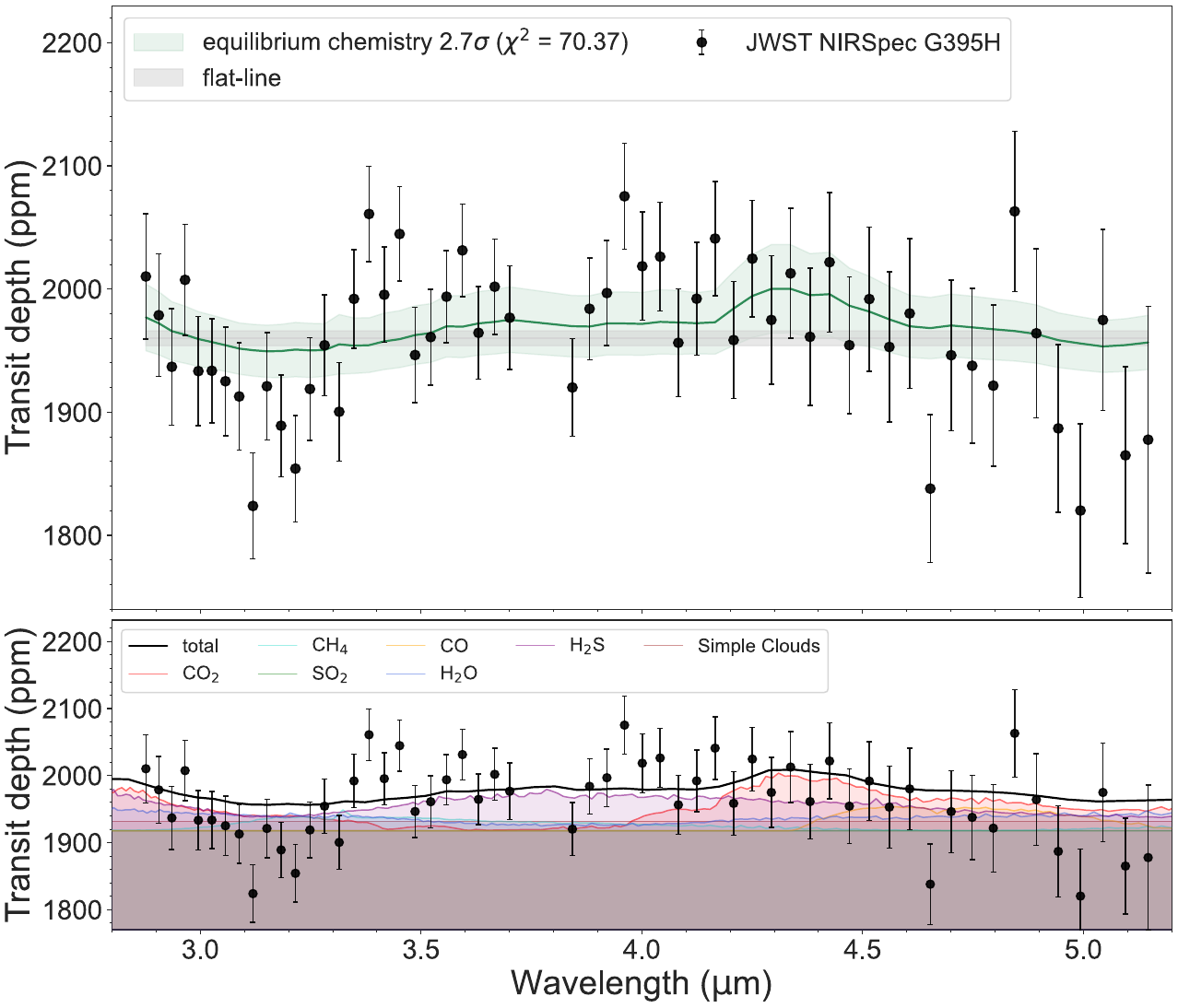}
    \caption{Top panel: Transmission spectrum of L98-59\,d obtained using \texttt{ transitspectroscopy} and binned to R $\sim$ 100 (black points) compared to an equilibrium chemistry retrieval model from \texttt{ TauREx} (light green). Bottom : Opacity contributions from the best-fit model. The model suggests a contribution from H$_2$S (purple) and CO$_2$ (red) opacities. }
    \label{fig:atmo_retrieval_eq}
\end{figure*}

\begin{figure*}
    \centering
  \includegraphics[width =18cm]{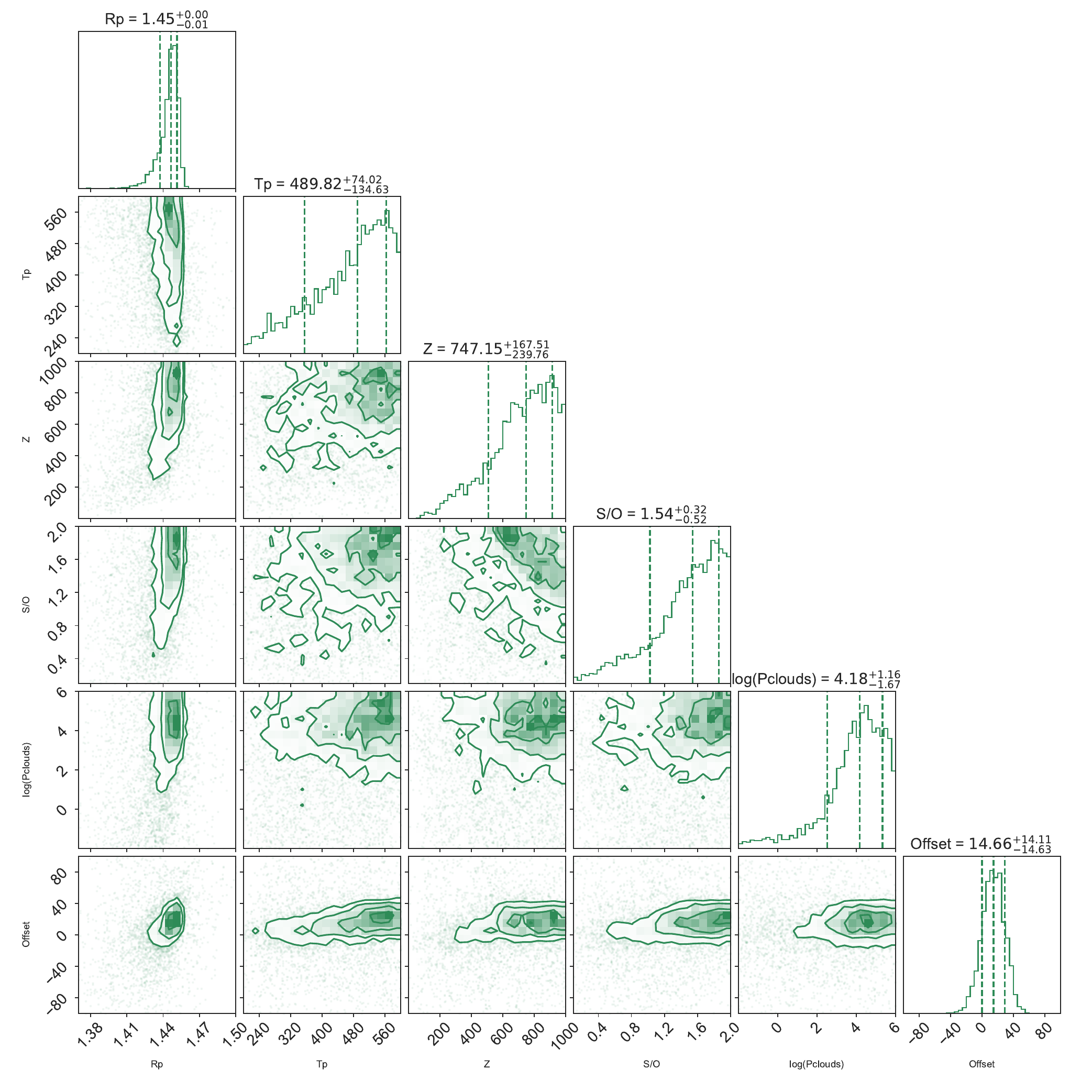}
    \caption{Posterior distributions from the equilibrium chemistry retrieval analysis. Not all parameters in the fit are plotted for clarity.  }
    \label{fig:atmo_retrieval_eq_post}
\end{figure*}

\subsection{Stellar contamination retrieval analysis}\label{Appendix:retrieval:stellar}
\begin{figure*}
    \centering
  \includegraphics[width =18cm]{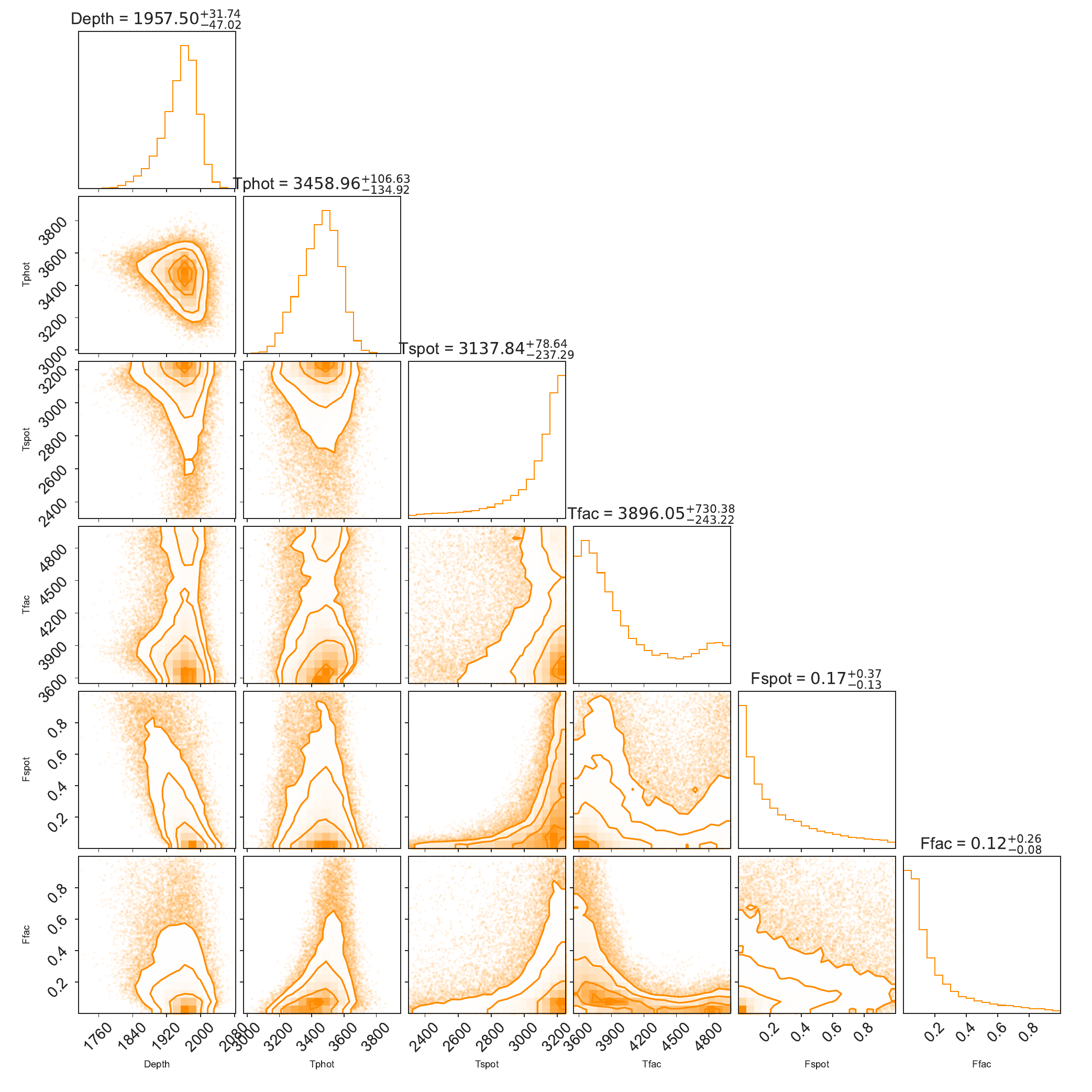}
    \caption{Posterior distributions from the stellar retrieval analysis. }
    \label{fig:stellar_retrieval_post}
\end{figure*}

\subsection{Atmospheric retrieval analysis on the \texttt{FIREFLy} data reduction}\label{Appendix:retrieval:firefly}
\begin{figure*}[htpb]
    \centering
  \includegraphics[width=\textwidth]{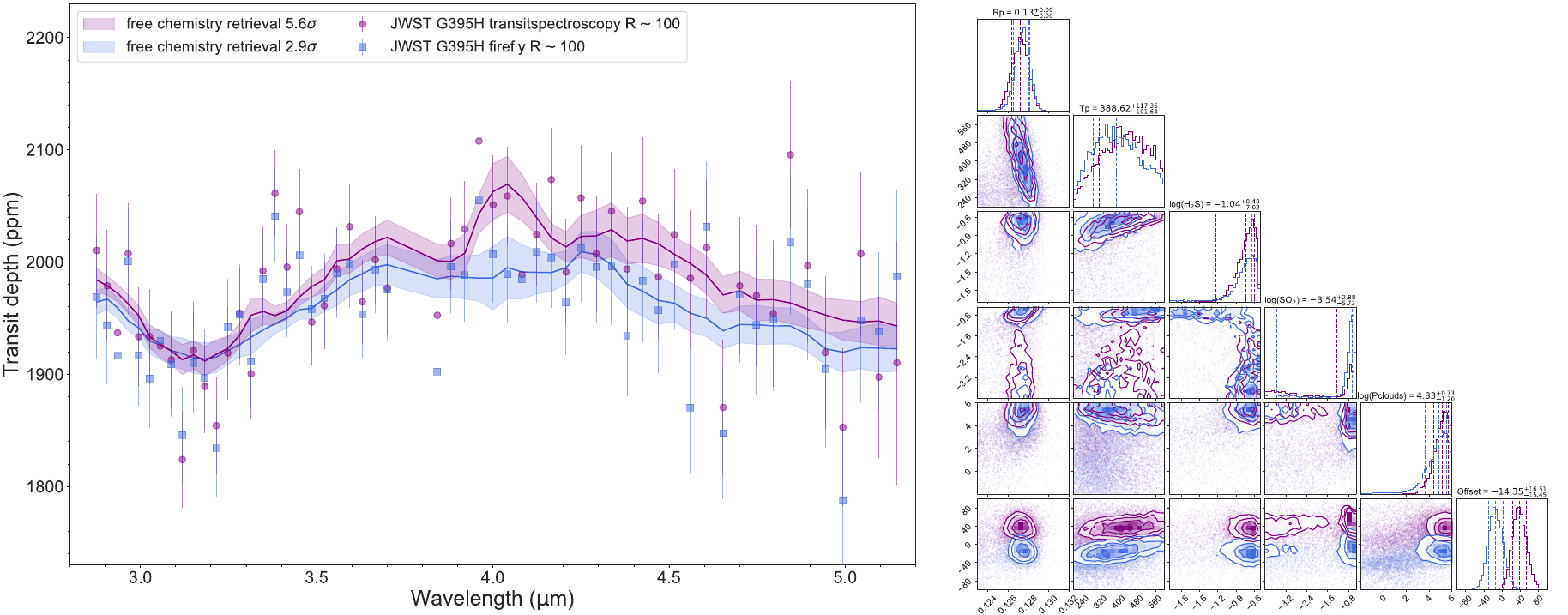}

  \includegraphics[width=\textwidth]{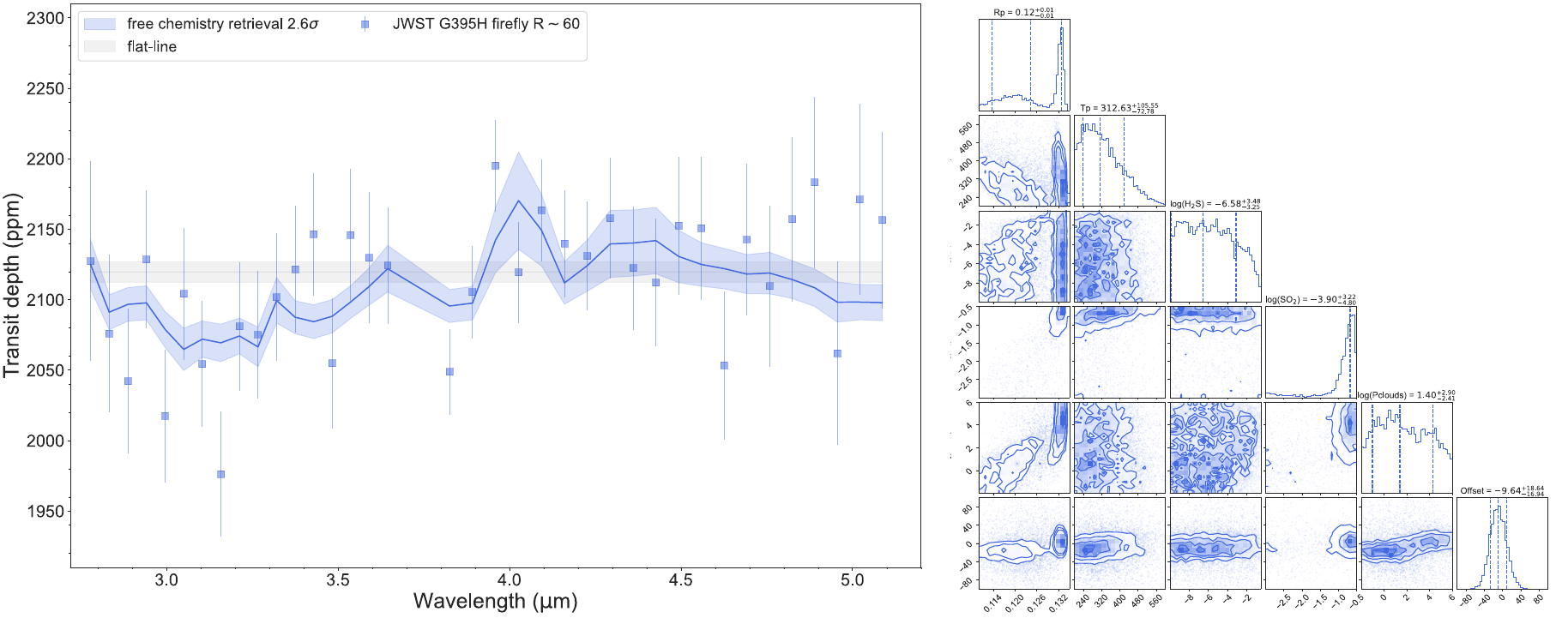}
    \caption{Top Left:Transmission spectrum of L98-59\,d obtained using \texttt{ transitspectroscopy} (pink data points) and  \texttt{FIREFLy}  (blue data points) and binned to R $\sim$ 100 compared to atmospheric retrievals models with 1$\sigma$ uncertainty. An Offset is fitted between NRS1 and NRS2. Top Right : Posterior distributions from the atmospheric retrieval analyses. We do not represent all parameters for clarity. The values are printed for the  \texttt{FIREFLy}  atmospheric fit. 
    Bottom Left :Transmission spectrum of L98-59\,d obtained using  \texttt{FIREFLy}  (blue data points) binned to R $\sim$ 60 compared to an atmospheric retrieval models with 1$\sigma$ uncertainty. An Offset is fitted between NRS1 and NRS2. Bottom Right : Posterior distributions from the atmospheric retrieval analysis. }
    \label{fig:retrieval_comparison}
\end{figure*}

\subsection{Does the offset between NRS1 and NRS2 impact the atmospheric detection ?} \label{Appendix:retrieval:offset}
To assess the influence of introducing an offset between the NRS1 and NRS2 transit depth values, we conducted an atmospheric retrieval fit on the transmission spectrum of L98-59\,d at R $\sim$ 100 obtained from \texttt{ transitspectroscopy}. The best-fit model is shown in Figure\,\ref{fig:atmo_retrieval} bottom panel along with the stellar retrieval model. The atmospheric model is preferred at 3.4$\sigma$ compared to a flat-line with SO$_2$ and H$_2$S as the main opacities. Moreover, incorporating these two molecules is favored at a significance of 3.2$\sigma$ compared to a model without them. Introducing the offset led to higher NRS2 transit depths in \texttt{ transitspectroscopy}'s data, thereby reinforcing the atmospheric detection, particularly of SO$_2$ at 4.2$\upmu$m. However, removing the offset did not impact the detection of an atmosphere in \texttt{ transitspectroscopy}'s reduction. 

\subsection{Does the spectral resolution impact the atmospheric detection ?} \label{Appendix:retrieval:resolution}
To examine the impact of spectral resolution on our results, we applied the atmospheric free chemistry retrieval setup to \texttt{ transitspectroscopy}'s data binned at different resolutions, following the fit at pixel-level resolution. We then evaluated the atmospheric detection by comparing Bayesian evidences with a flat-line, representing a fit without opacity sources. We did not fit for an offset between NRS1 and NRS2 in this analysis. The choice of spectral resolution did not significantly affect the atmospheric detection obtained with \texttt{ transitspectroscopy}'s reduction.

\end{document}